\documentclass[runningheads]{llncs}
\usepackage{graphicx}

\begin{document}
\title{Evaluating the Performance of StyleGAN2-ADA on Medical Images}

\titlerunning{StyleGAN2-ADA on Medical Images}

\author{
McKell Woodland\inst{1,2}\and 
John Wood\inst{1} \and
Brian M. Anderson\inst{1,4} \and 
Suprateek Kundu\inst{1}
Ethan Lin\inst{1} \and
Eugene Koay\inst{1} \and
Bruno Odisio\inst{1} \and
Caroline Chung\inst{1} \and
Hyunseon Christine Kang\inst{1} \and 
Aradhana M. Venkatesan\inst{1} \and
Sireesha Yedururi\inst{1} \and
Brian De\inst{1} \and 
Yuan-Mao Lin\inst{1} \and 
Ankit B. Patel\inst{2,3} \and
Kristy K. Brock\inst{1} 
}

\authorrunning{Woodland et al.}
%
\institute{The University of Texas MD Anderson Cancer Center, Houston TX 77030, USA \email{MEWoodland@mdanderson.org}
\and
Rice University, Houston TX 77005, USA\and
Baylor College of Medicine, Houston TX 77030, USA\and
University of California San Diego, La Jolla CA 92093, USA}

\maketitle              

\begin{abstract}
Although generative adversarial networks (GANs) have shown promise in medical imaging, they have four main limitations that impede their utility: computational cost, data requirements, reliable evaluation measures, and training complexity.
Our work investigates each of these obstacles in a novel application of StyleGAN2-ADA to high-resolution medical imaging datasets.
Our dataset is comprised of liver-containing axial slices from non-contrast and contrast-enhanced computed tomography (CT) scans.
Additionally, we utilized four public datasets composed of various imaging modalities.
We trained a StyleGAN2 network with transfer learning (from the Flickr-Faces-HQ dataset) and data augmentation (horizontal flipping and adaptive discriminator augmentation).
The network’s generative quality was measured quantitatively with the Fr\'{e}chet Inception Distance (FID) and qualitatively with a visual Turing test given to seven radiologists and radiation oncologists. 

The StyleGAN2-ADA network achieved a FID of 5.22 ($\pm$ 0.17) on our liver CT dataset.
It also set new record FIDs of 10.78, 3.52, 21.17, and 5.39 on the publicly available SLIVER07, ChestX-ray14, ACDC, and Medical Segmentation Decathlon (brain tumors) datasets.
In the visual Turing test, the clinicians rated generated images as real 42\% of the time, approaching random guessing. 
Our computational ablation study revealed that transfer learning and data augmentation stabilize training and improve the perceptual quality of the generated images.
We observed the FID to be consistent with human perceptual evaluation of medical images.
Finally, our work found that StyleGAN2-ADA consistently produces high-quality results without hyperparameter searches or retraining.


\end{abstract}

\section{Introduction}

Recently, generative adversarial networks (GANs) have shown promise in many medical imaging tasks, including data augmentation in computer-aided diagnosis \cite{Pang2021}, image segmentation \cite{Xun2022}, image reconstruction \cite{Luo2021}, treatment planning \cite{Aleef2021}, image translation \cite{Jiang2021}, and anomaly detection \cite{Schlegl2017}.
Despite their potential in medical imaging, GANs have several drawbacks that impede both their capabilities and utilization in the medical field.
These obstacles include computational cost, data requirements, flawed measures of assessment, and training complexity.

GANs are computationally expensive.
The original StyleGAN2 project took 51.06 GPU years to create, 0.23 of which were used for training the Flickr-Faces-HQ (FFHQ) weights used in our paper \cite{Karras2020_2}.
Despite being the state-of-the-art generative model for high-resolution images, StyleGAN2 is often not used in medical imaging literature due to its expense \cite{Segal2021}.
If it is used, images are brought to lower resolutions to offset the cost \cite{Montero2021,Pocev2021}.
While StyleGAN \cite{Karras2019} (the predecessor to StyleGAN2) has been applied to high-resolution medical images \cite{Fetty2020}, we believe our paper is the first rigorous evaluation of StyleGAN2 on multiple high-resolution medical imaging datasets.

At high-resolutions, GANs require hundreds of thousands of images to effectively train, a requirement that is extremely challenging to satisfy in the medical field.
With limited data, the GAN's discriminator overfits on the training examples, obstructing the GAN's ability to converge.
Adaptive discriminator augmentation (ADA) was designed to reduce discriminator overfitting through a wide range of data augmentations that do not ``leak'' to the generated distribution.
When applied to a histopathology dataset, ADA improved the FID by 84\% \cite{Karras2020_ada}.
In our paper, we perform a computational ablation study that examines how ADA and transfer learning affects performance on medical images.

One of the greatest challenges in GANs is constructing robust quantitative evaluation measures \cite{Lucic2018}.
The Fr\'{e}chet Inception Distance (FID) \cite{Heusel2017} is the standard for state of the art evaluation for generative modeling in natural imaging.
It relies on an Inception network that was trained on ImageNet, which does not contain medical images \cite{deng2009}, for its calculation.
As such, a common assumption in related literature is that the FID is not applicable to medical images.
We revisit this assumption by testing the correlation between the FID and human perceptual evaluation on medical images.

GANs are notoriously challenging to train.
They have numerous hyperparameters and suffer from training instability.
In a large empirical evaluation of various GANs, Lučić et al. \cite{Lucic2018} found that GAN training is extremely sensitive to hyperparameter settings.
A separate study illustrated this sensitivity by performing 1,500 hyperparameter searches on three unique medical imaging datasets with various GAN architectures.
The authors found that few models produced meaningful images; even fewer models achieved reasonable metric evaluations \cite{skandarani2021}.
Neither of these studies examined StyleGAN2.
Our work is unique in that we test the stability of StyleGAN2, along with its ability to generate quality images without a hyperparameter search.

The main contributions of our research are as follows:
\begin{itemize}
    \item We apply StyleGAN2 to a variety of high-resolution medical imaging datasets.
    \item We perform a computational ablation study on the effect of transfer learning and data augmentation on a limited-data medical imaging dataset.
    \item We provide empirical evidence that the FID is consistent with human perceptual evaluation of medical images.
    \item We evaluate StyleGAN2's stability and ability to produce quality results without a hyperparameter search.
    \item We achieve state-of-the-art FIDs on four public datasets.
\end{itemize}

\section{Methods}

\subsection{Data}

We used the 97 non-contrast and 108 contrast enhanced abdominal computed tomography (CT) scans presented in \cite{Anderson2021}.
To accentuate the liver, the data was windowed to a level 50 and a width 350, consistent with the preset values for viewing the liver in a commercial treatment planning system (RayStation v10, RaySearch Laboratories, Stockholm, Sweden).
All axial slices that contained no liver information were discarded.
Voxel values were mapped to the range $[0, 255]$ and converted each axial slice to a PNG image.
In all, our training dataset contained 10,600 512x512 images.
Three randomly sampled images from our training dataset are shown in the first row of Figure \ref{fig:images}.
We used an additional 143,345 512x512 images for one experiment in our ablation study.
These images were obtained by applying the above mentioned preprocessing steps to 3,029 abdominal CT scans (301 patients) that were retrospectively acquired under an IRB approved protocol.

Separately, our methods were applied to several publicly available datasets.
For the ``Segmentation of the Liver Competition 2007'' (SLIVER07) dataset\footnote{\url{https://sliver07.grand-challenge.org/}} \cite{Heimann2009}, we used the 20 scans available in the training dataset and converted each slice to a PNG image without any further preprocessing.
In total, this dataset consisted of 4,159 512x512 images.
To our knowledge, the previous best FID (29.06) on this dataset was achieved by Skandarani et al. using the StyleGAN network.

The ChestX-ray14 dataset\footnote{\url{https://nihcc.app.box.com/v/ChestXray-NIHCC}} \cite{wang2017chestxray} consists of 112,120 1024x1024 Chest X-ray images in PNG format.
The previous best FID on the ChestX-ray14 dataset of 8.02 was achieved using a Progressive Growing GAN \cite{Segal2021}. 
No preprocessing on this dataset was performed.
The Automated Cardiac Diagnosis Challenge (ACDC) dataset\footnote{\url{https://acdc.creatis.insa-lyon.fr/}} \cite{Bernard2018} consists of 150 cardiac cine-magnetic resonance imaging (MRI) exams.
We used the training dataset, which consists of 100 exams.
The images were rescaled to the range [0, 255] using SimpleITK \cite{Lowekamp2013} and padded with zeros.
Each slice was then converted to a 2D PNG image.
In total, this dataset consisted of 1,902 512x512 images.
The previous best FID on the ACDC training dataset (24.74) was achieved with StyleGAN \cite{skandarani2021}.

Additionally, we applied StyleGAN2-ADA to a dataset whose FID had not been previously evaluated: the brain tumor data from the Medical Segmentation Decathlon\footnote{\url{http://medicaldecathlon.com/}} \cite{Simpson2019}, which contains 750 4D MRI volumes.
The gadolinium-enhanced T1-weighted 3D images were extracted and windowed to the range [0, 255] using SimpleITK.
Slices were converted to 2D PNG images.
This dataset consists of 103,030 256x256 PNG images.

\subsection{Generative Modeling}

Due to its state-of-the-art performance on high-resolution images, we used a StyleGAN2 network as our generative model \cite{Karras2020_2}.
For our experiments, we utilized the StyleGAN2 configuration of the official StyleGAN3 repository\footnote{\url{https://github.com/NVlabs/stylegan3}} \cite{Karras2021}.
We used the default parameters provided by the implementation, with the exception of changing $\beta_0$ to 0.9 in the Adam optimizer and disabling mixed precision.
We did not perform a hyperparameter search.
We explored the effects of transfer learning and data augmentation in an ablation study with the following experimental designs:
\begin{enumerate}
    \item \textbf{Baseline} Disable all StyleGAN2 augmentations and train from scratch.
    \item \textbf{Pretrained} Disable all augmentations and begin training with pretrained weights from StyleGAN2 trained on the FFHQ dataset.
    \item \textbf{Augmented} Enable mirroring (horizontal flipping) and ADA and train from scratch.
    \item \textbf{Pretrained and Augmented} Enable mirroring and ADA and begin training with the official FFHQ StyleGAN2 weights.
\end{enumerate}
Each of these experiments was performed on our liver CT training dataset.
A variation of Experiment 1 was also performed where 143,345 liver images were added to the training dataset.
Furthermore, Experiment 4 was performed on the four public datasets.
Each experiment was performed on a DGX with eight 40GB A100 GPUs.
DGXs were accessed using the XNAT platform \cite{marcus2007}.
Experiments ran for 6,250 ticks with metrics calculated and weights saved every 50 ticks.
Each experiment took approximately 1.5, 4, and 7 days to complete for 256x256, 512x512, and 1024x1024 sized datasets, respectively.
We repeated each experiment five times to test algorithm stability.

\subsection{Evaluation Measures}

\subsubsection{Fr\'{e}chet Inception Distance}

The FID is the standard for state of the art GAN evaluation in natural imaging.
It is the Fr\'{e}chet distance between two multivariate Gaussians constructed from representations extracted from the coding layer of an Inception network that was pretrained on ImageNet \cite{Heusel2017}.
Several advantages of the FID include its ability to distinguish generated from real samples, agreement with human perceptual judgements, sensitivity to distortions, and computational and sample efficiency \cite{borji2019,Heusel2017}.
As such, we used the FID as our quantitative metric.
For each run, we reported the best FID achieved during training.
We used the model weights associated with each best FID for further qualitative analysis.
For statistical testing, we used permutation tests with $\alpha=0.05$.

Because ImageNet does not contain medical images, prior publications have argued that the FID is not applicable to medical imaging \cite{Chen2021,jung2021eijin,Tronchin2021}.
As such, they substitute the Inception network with their own encoding networks.
This trend has several limitations.
First, the FID is only consistent as a metric inasmuch as the same encoding model is used.
By using a new model, the reported distance can no longer be considered in the context of prior work that utilizes the FID.
Second, the algorithm designer is formulating their own evaluation metric, which will likely introduce unquantified bias into the presented results.
Due to these limitations, we use the original definition of the FID for our calculations.

\subsubsection{Visual Turing Tests}

Because the applicability of the FID to medical imaging is not well understood, our first visual Turing test evaluated the correlation between the FID and human perception on medical images.
The test was administered in a Google Form with four sections (created in random order), one per experiment.
Each section contained 40 randomly shuffled images, 20 real and 20 generated.
All images were randomly selected and only appeared once in the test.
The test was given to five participants with a medical physics background who were not familiar with the images.
We evaluated the test with the false positive rate (FPR) and false negative rate (FNR).

The purpose of the second visual Turing test was to rigorously validate the perceptual quality of the images generated by the pretrained StyleGAN2-ADA model on our dataset.
This test consisted of 50 real and 50 generated images randomly sampled and shuffled.
Each section contained one image, a question asking the participant if the image was real or fake, and a Likert scale assessing how realistic the image was.
The Likert scale was between 1 (fake) and 5 (real).
The test was given to seven radiologists or radiation oncologists with an average of 10 years of radiological experience.
The results of the Turing test were evaluated with precision, recall, accuracy, FPR, and FNR metrics.
Additionally, we computed the average Likert values for both real and generated images.
For statistical testing, we used permutation tests with $\alpha=0.10$.

\section{Results}

\begin{figure}
    \centering
    \begin{picture}(347,484)
        \put(25,400){\includegraphics[width=0.27\textwidth]{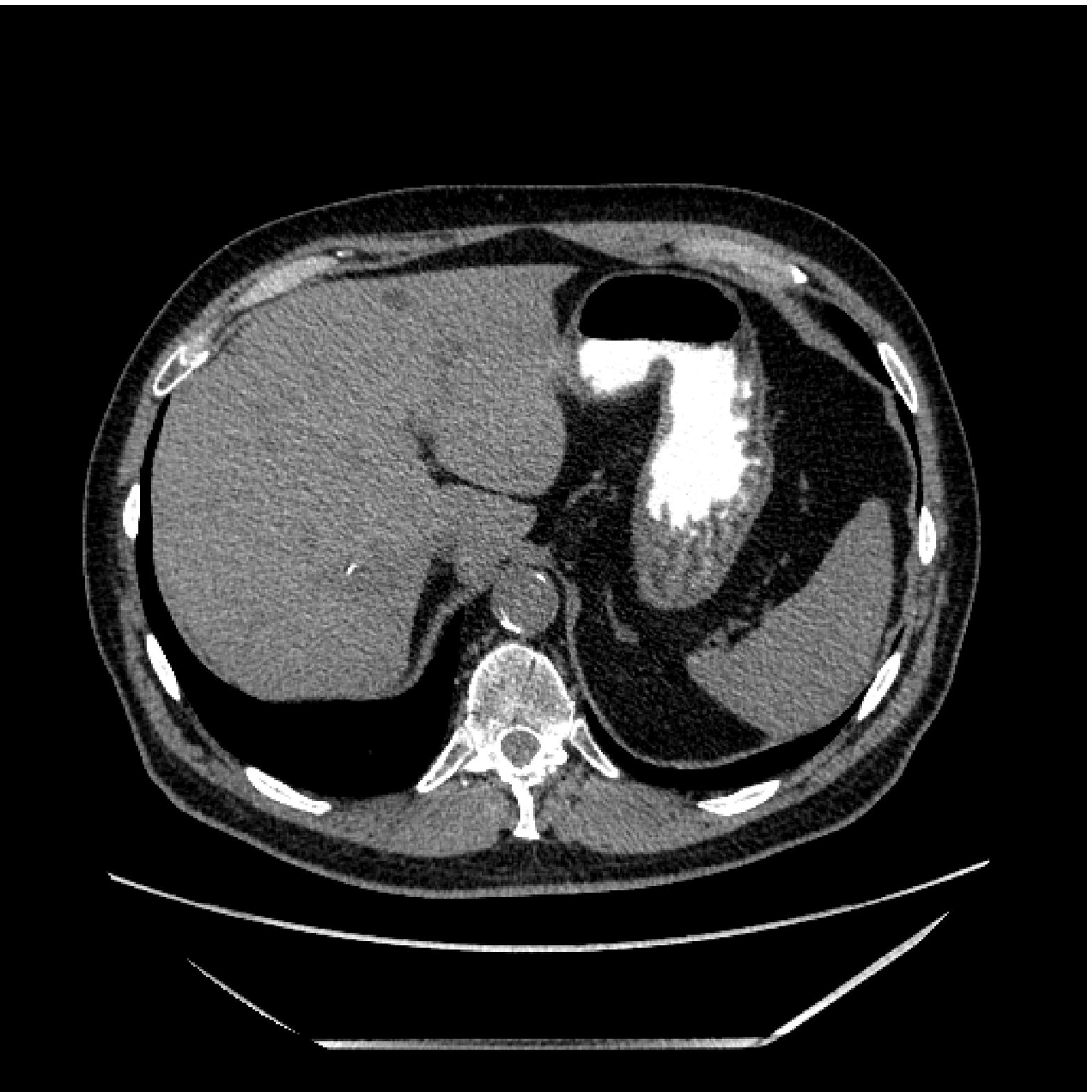}}
        \put(125,400){\includegraphics[width=0.27\textwidth]{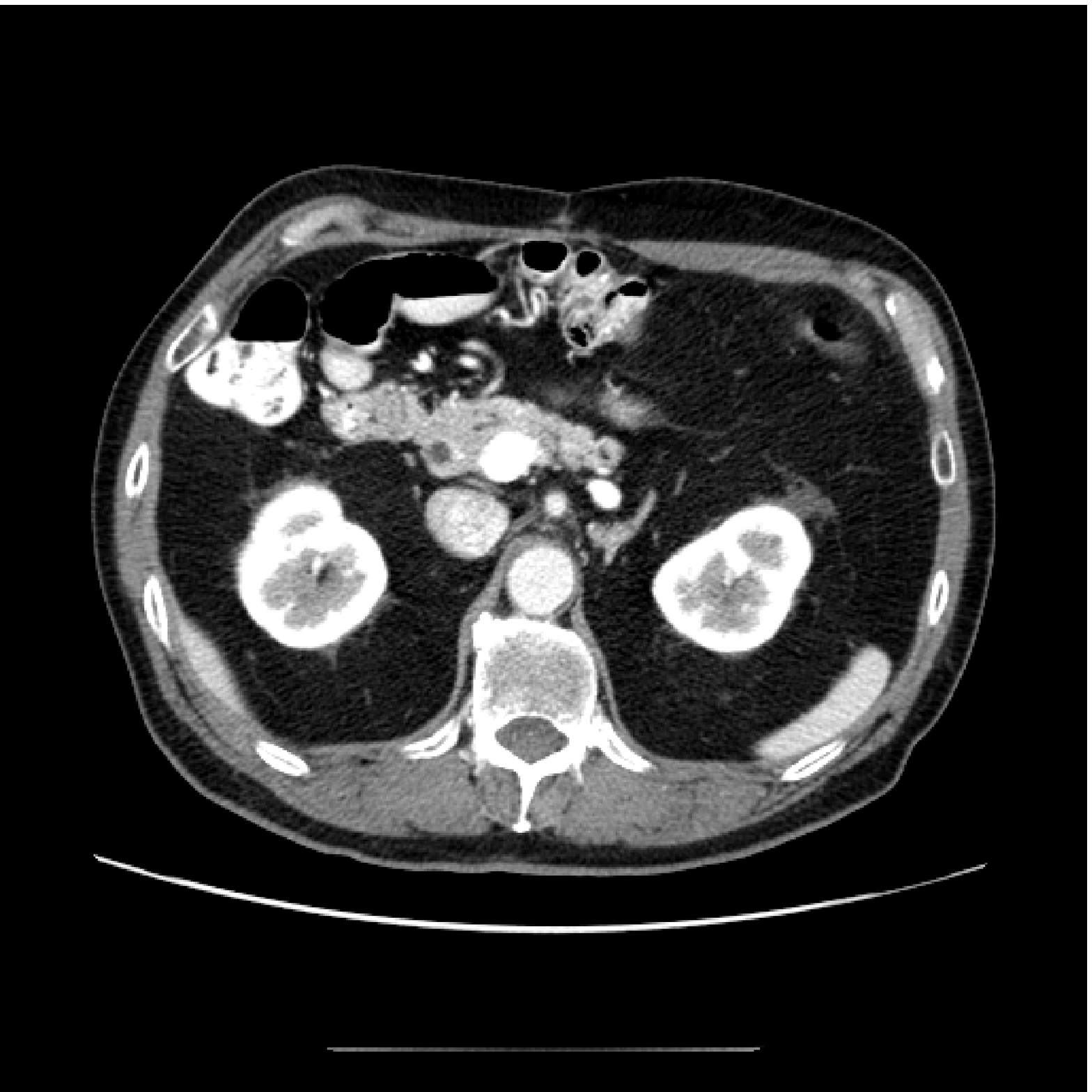}}
        \put(225,400){\includegraphics[width=0.27\textwidth]{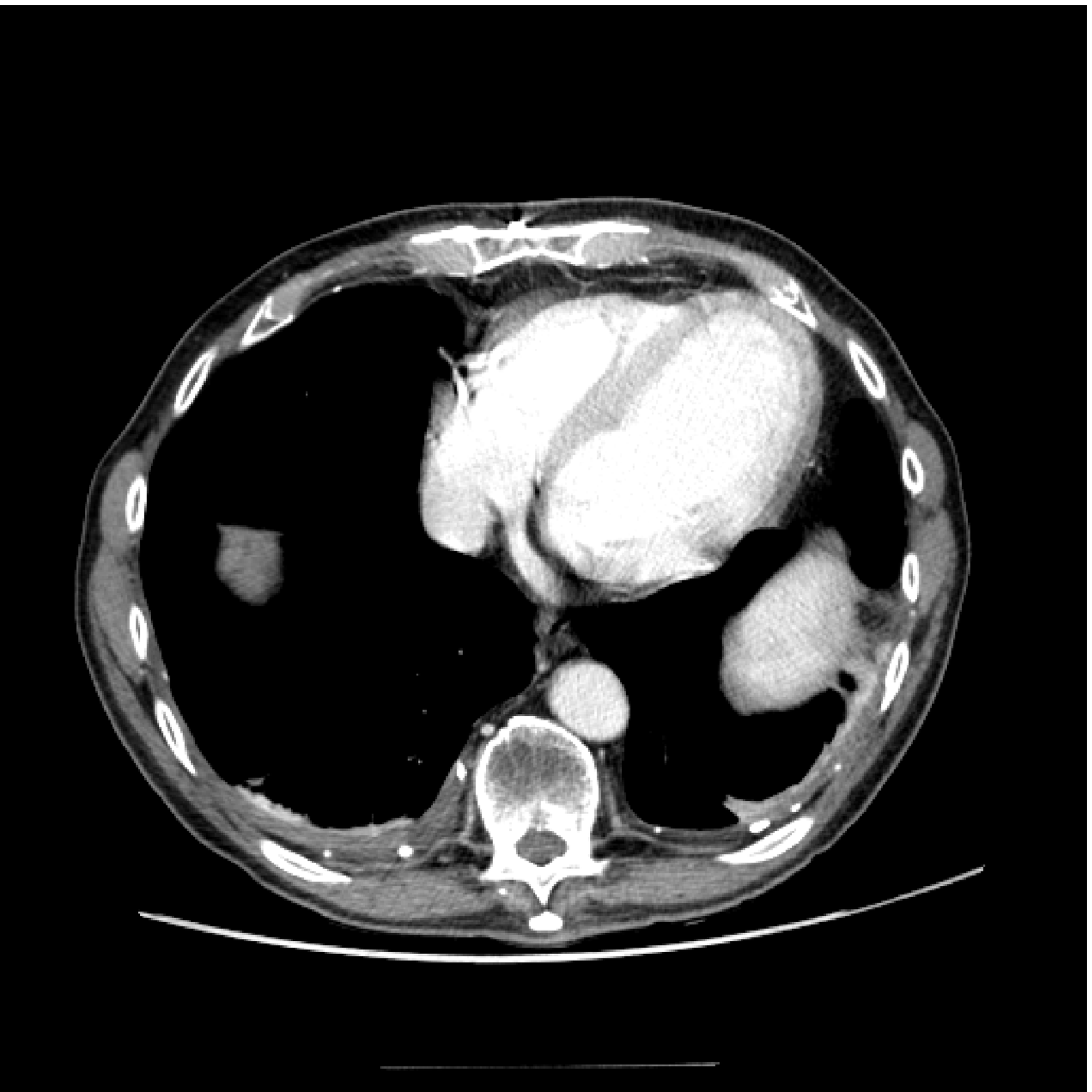}}
        \put(25,300){\includegraphics[width=0.27\textwidth]{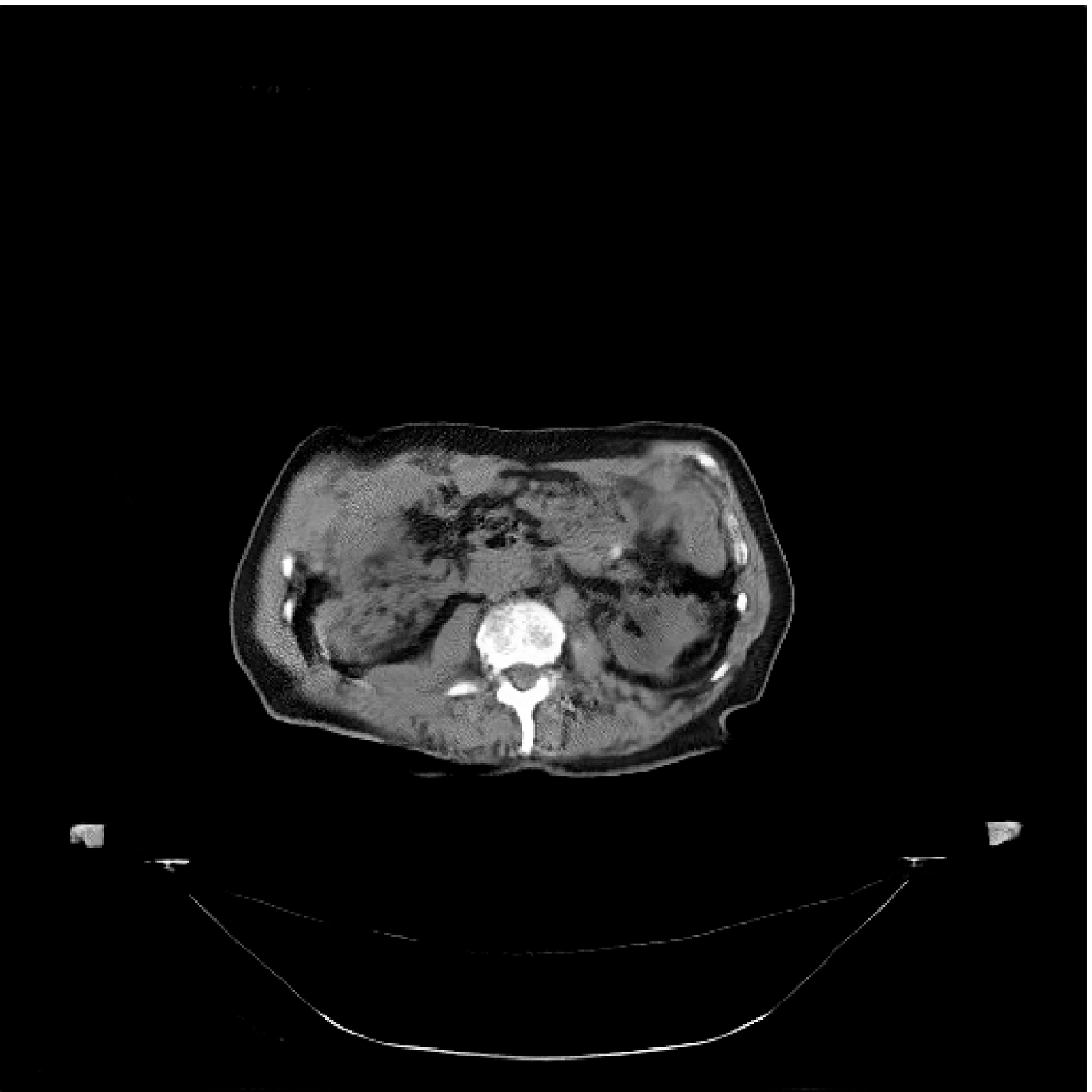}}
        \put(125,300){\includegraphics[width=0.27\textwidth]{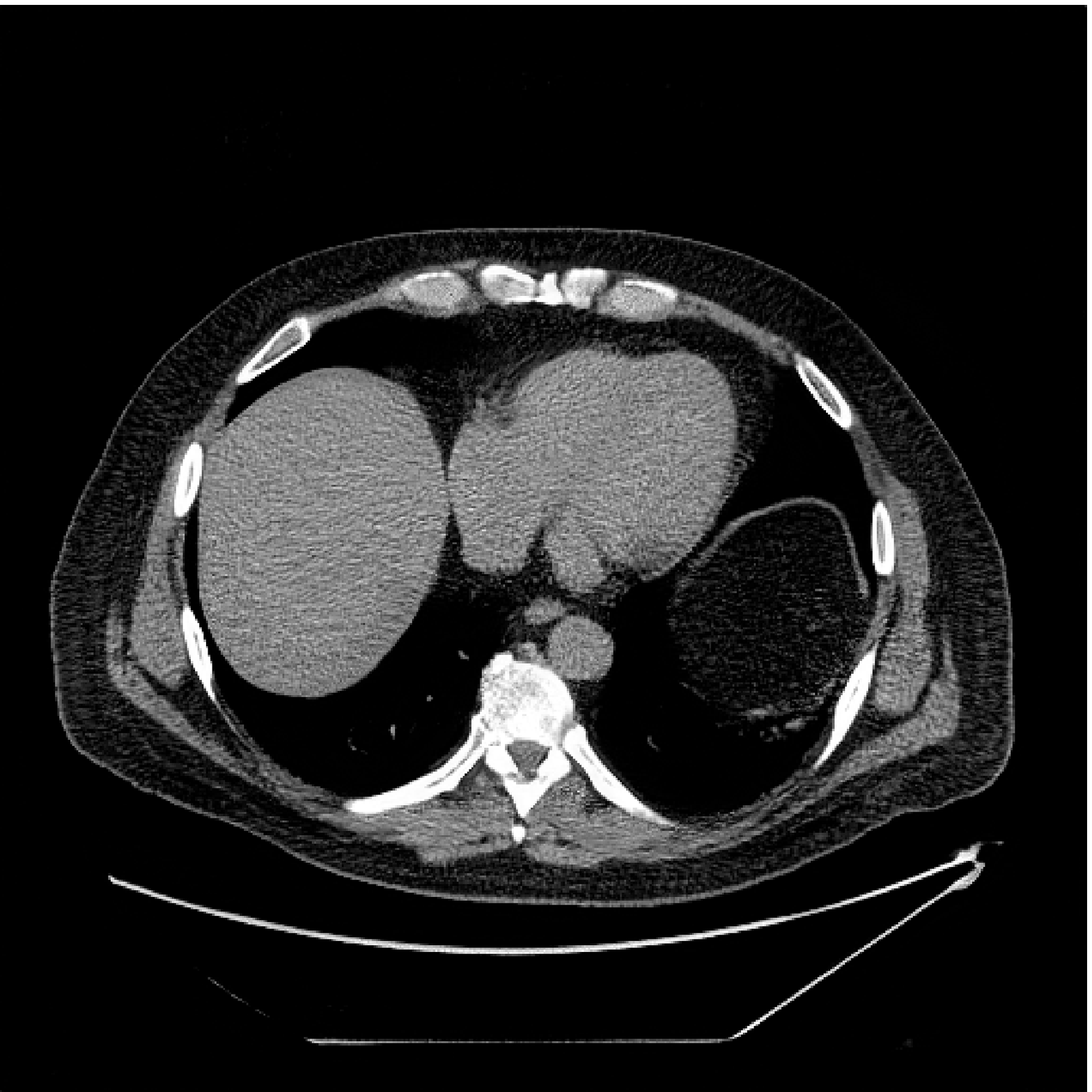}}
        \put(225,300){\includegraphics[width=0.27\textwidth]{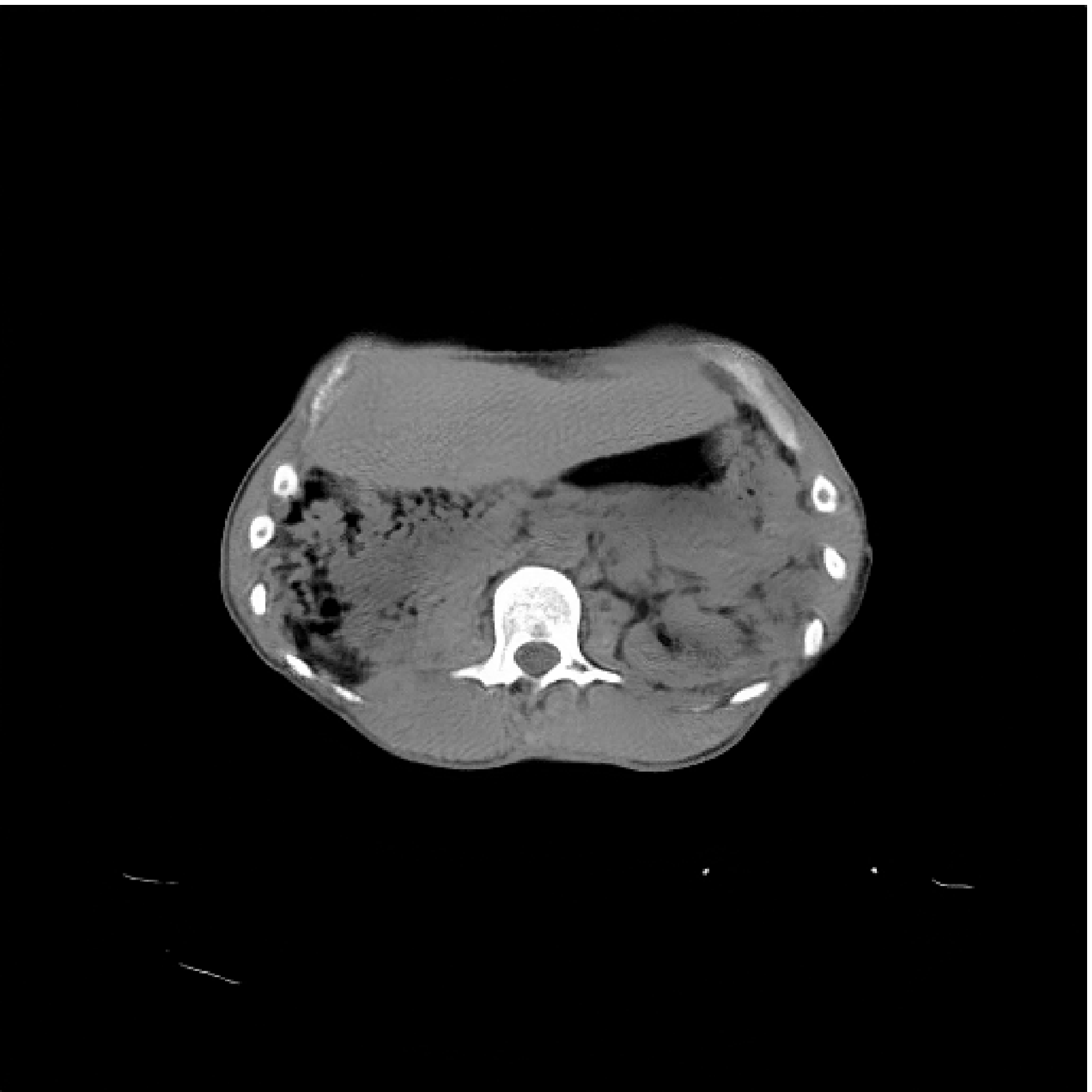}}
        \put(25,200){\includegraphics[width=0.27\textwidth]{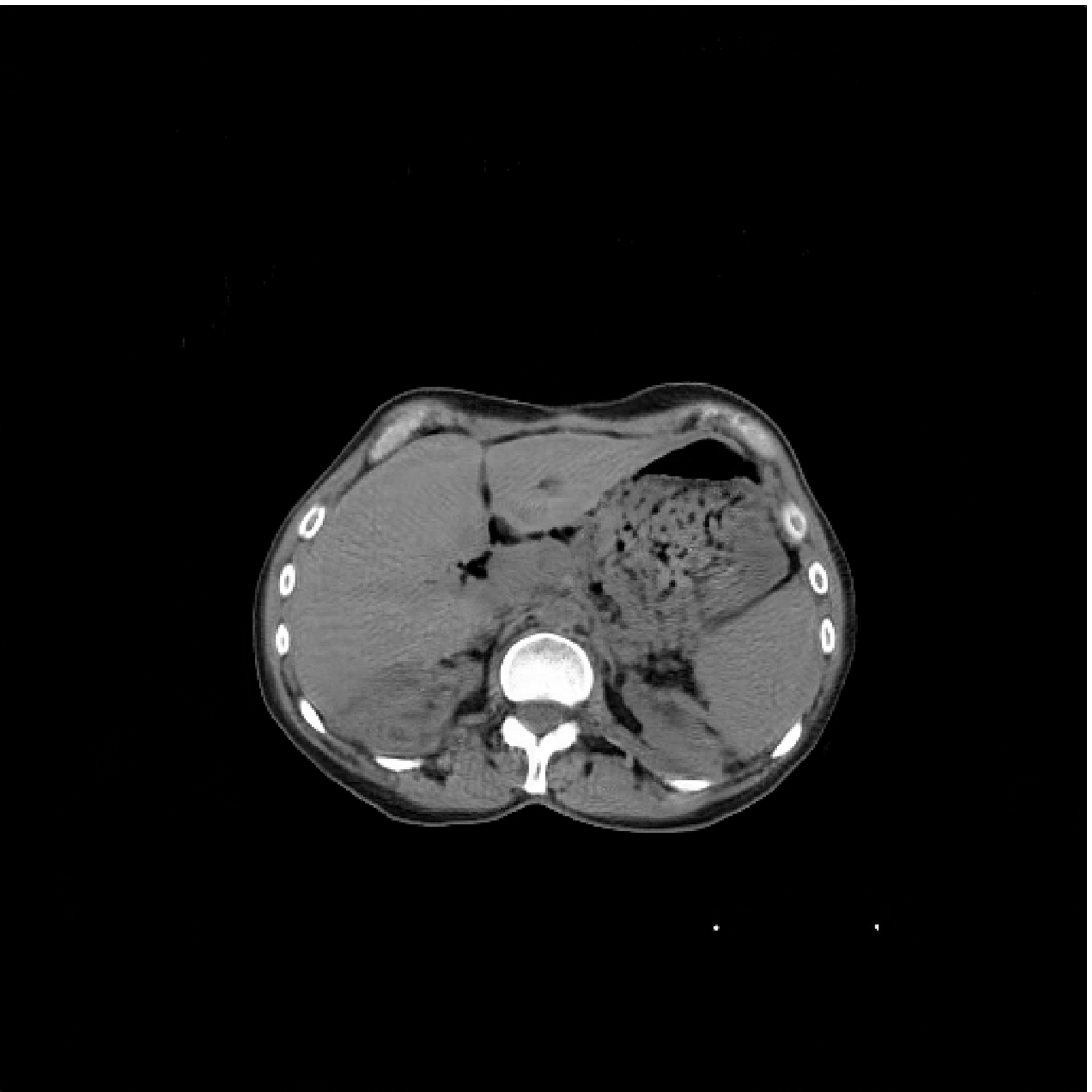}}
        \put(125,200){\includegraphics[width=0.27\textwidth]{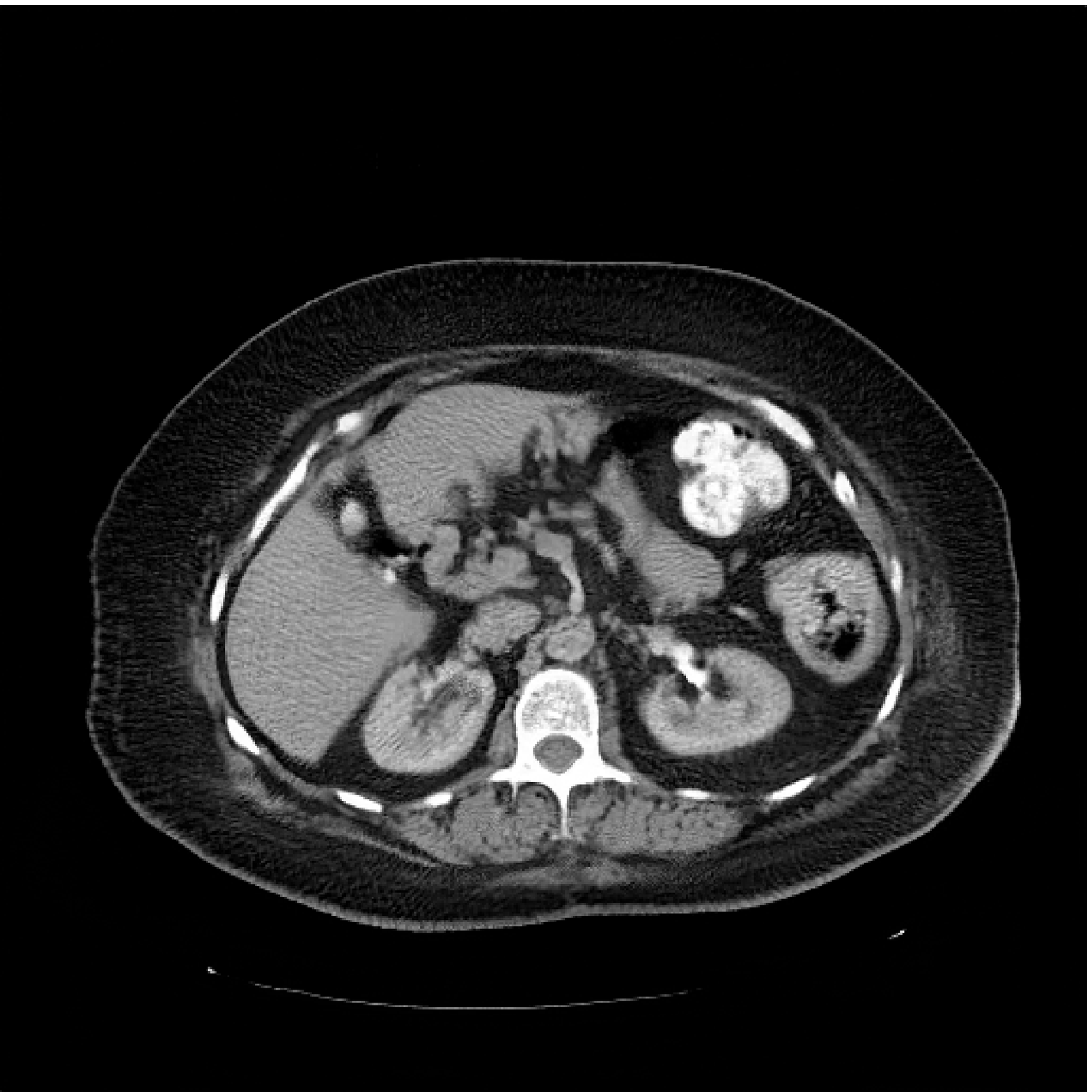}}
        \put(225,200){\includegraphics[width=0.27\textwidth]{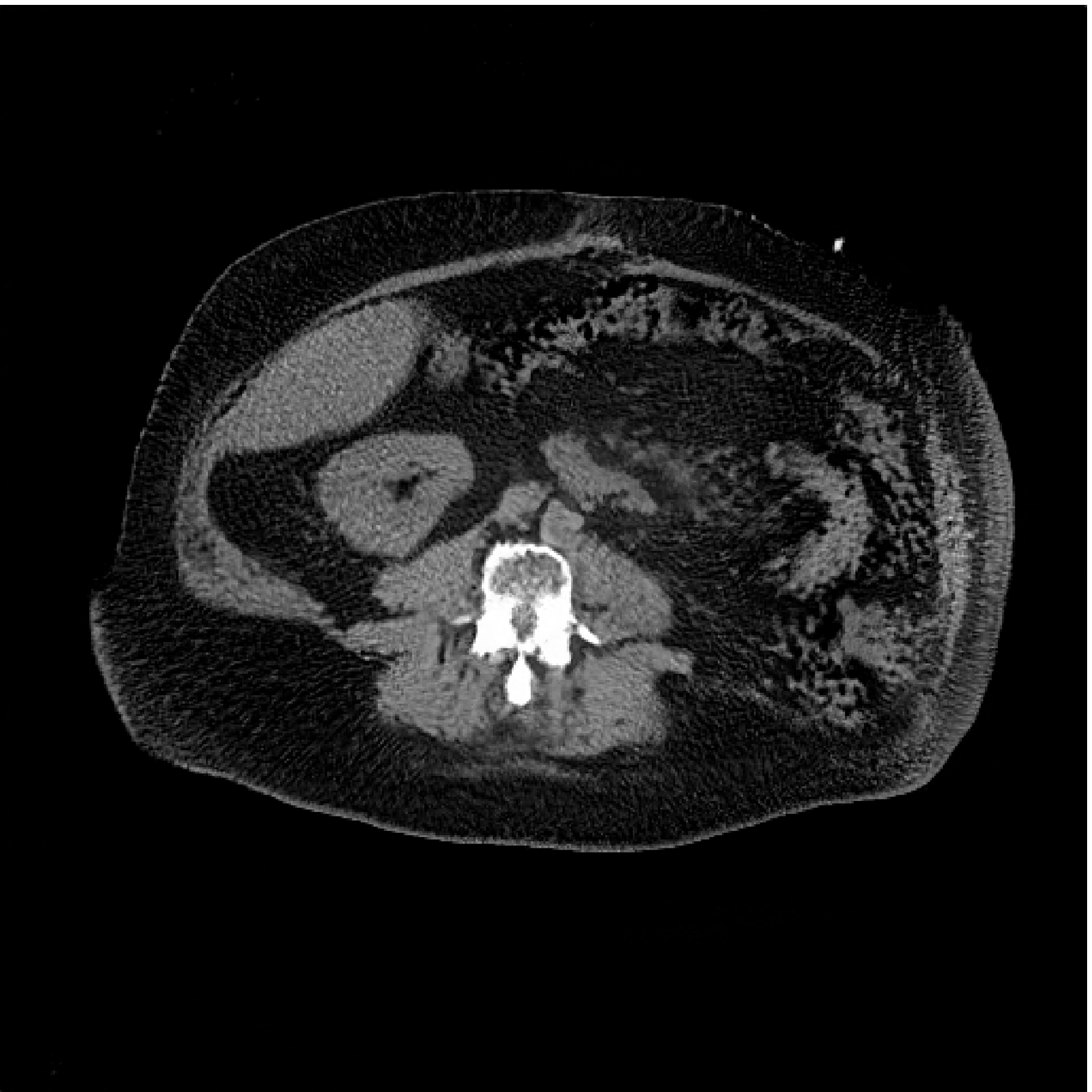}}
        \put(25,100){\includegraphics[width=0.27\textwidth]{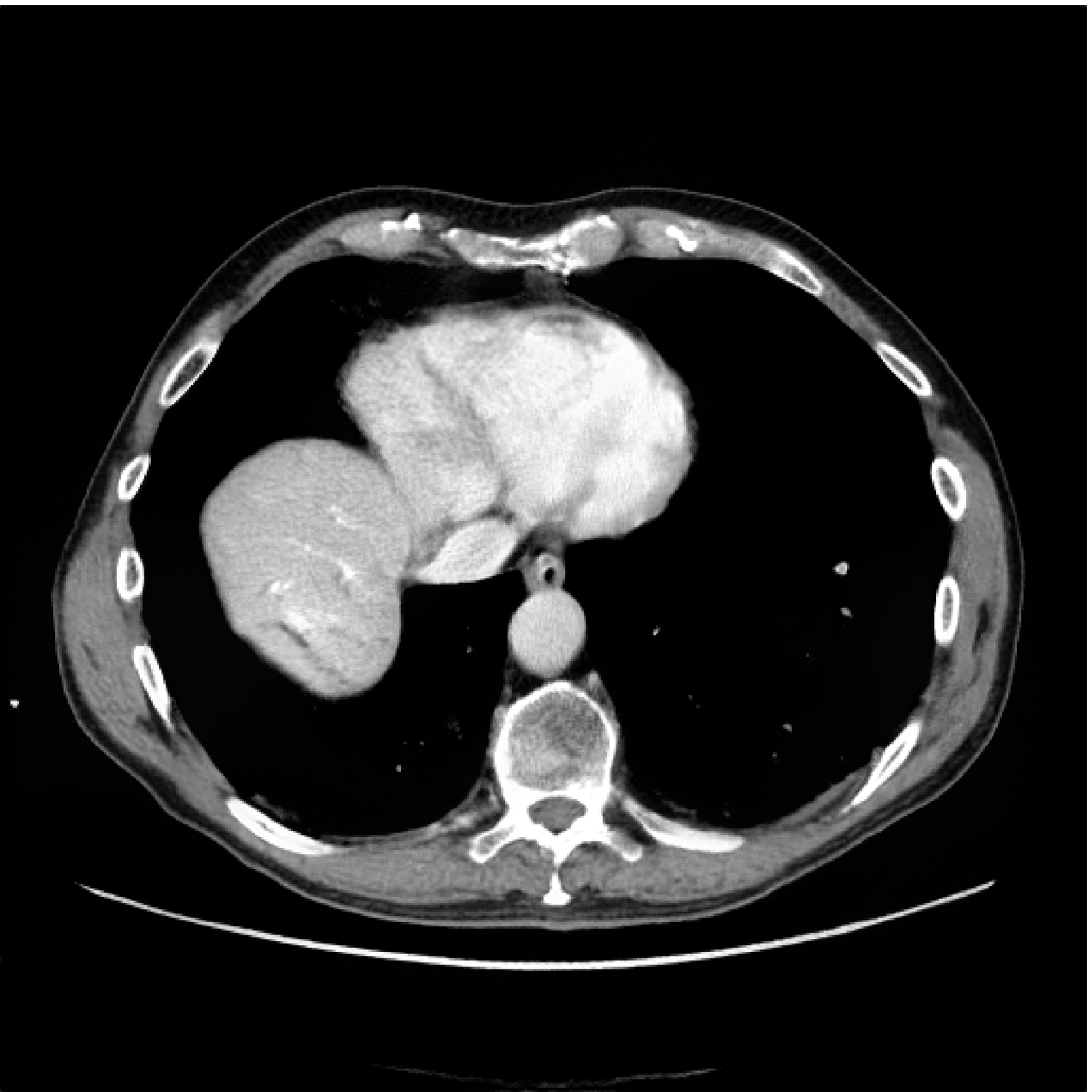}}
        \put(125,100){\includegraphics[width=0.27\textwidth]{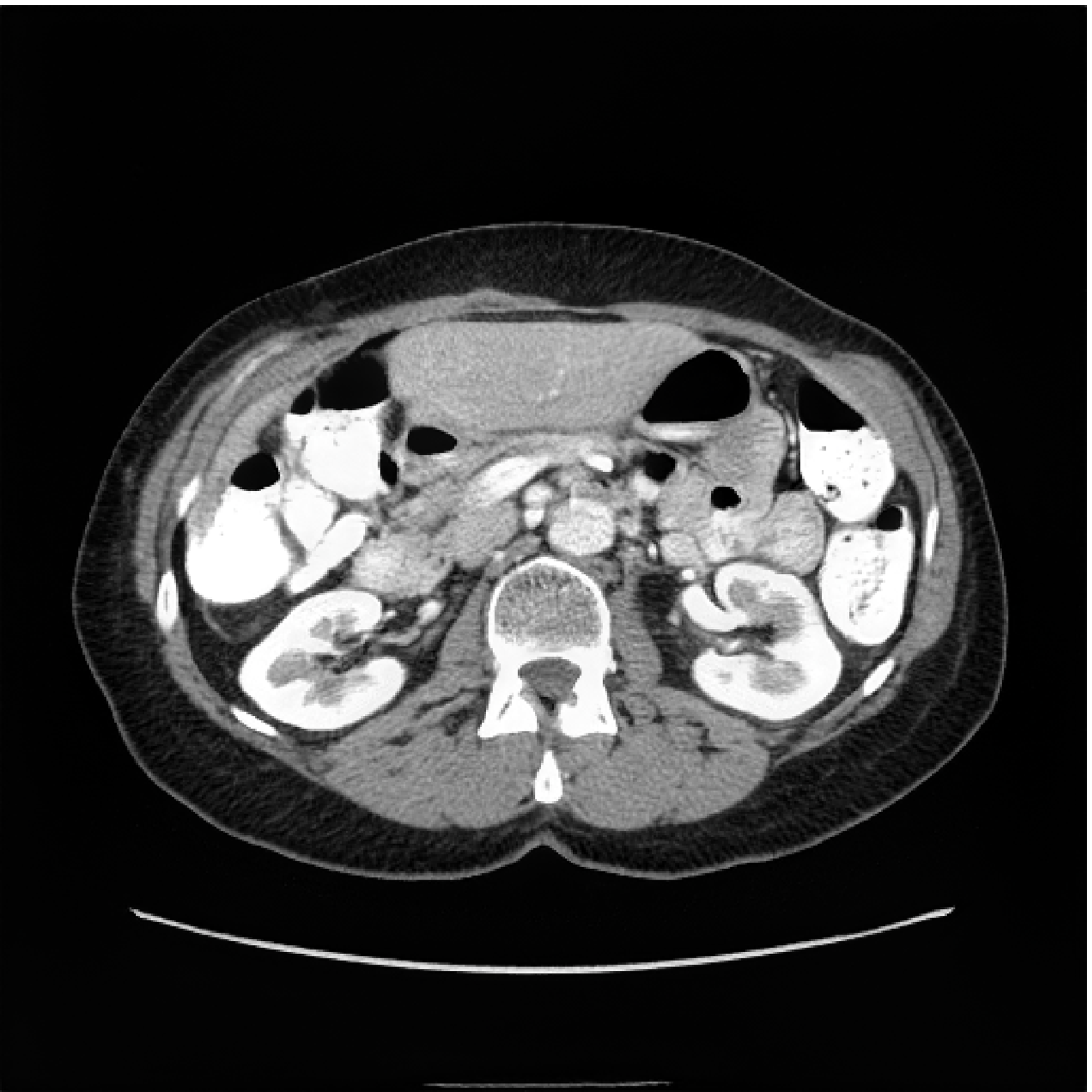}}
        \put(225,100){\includegraphics[width=0.27\textwidth]{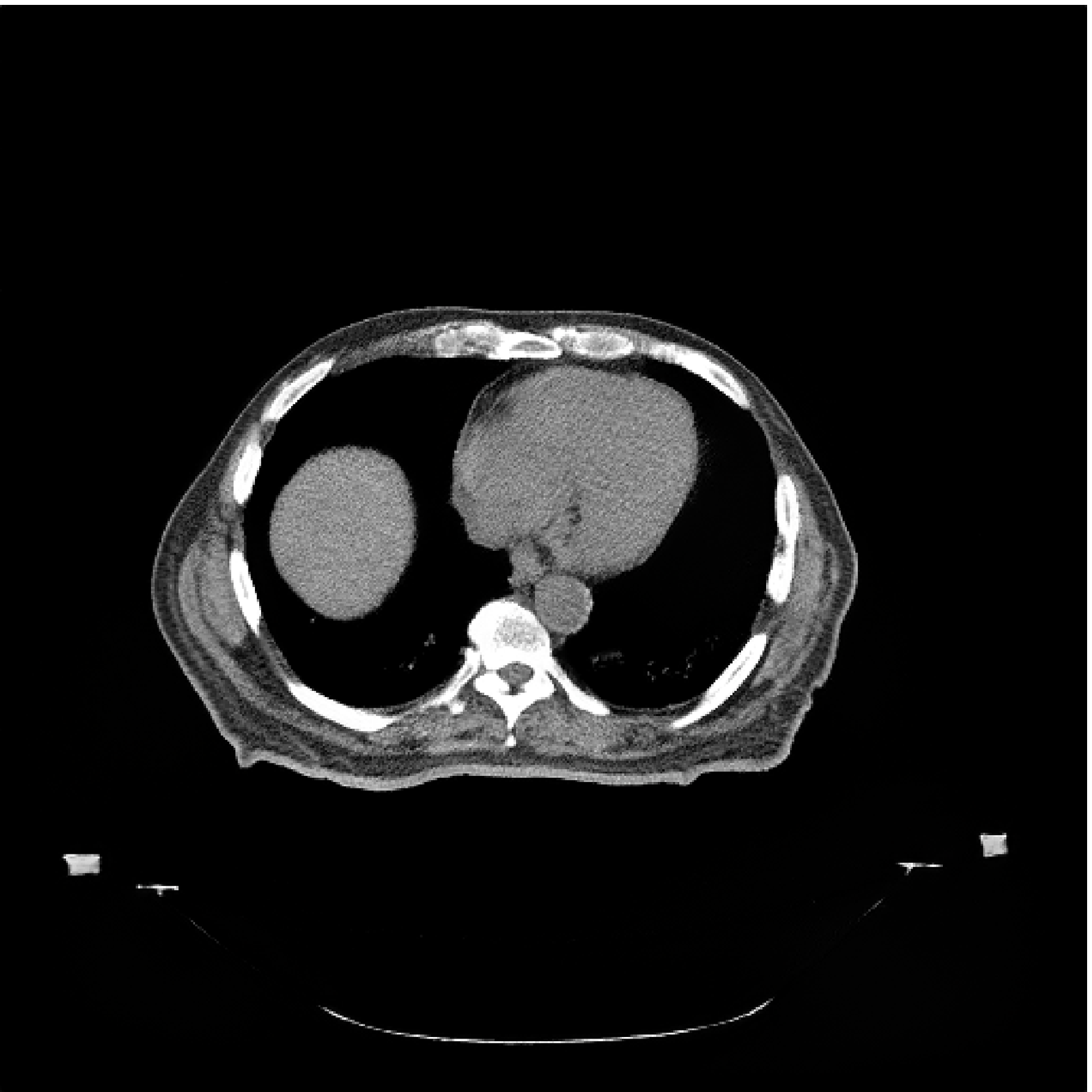}}
        \put(25,0){\includegraphics[width=0.27\textwidth]{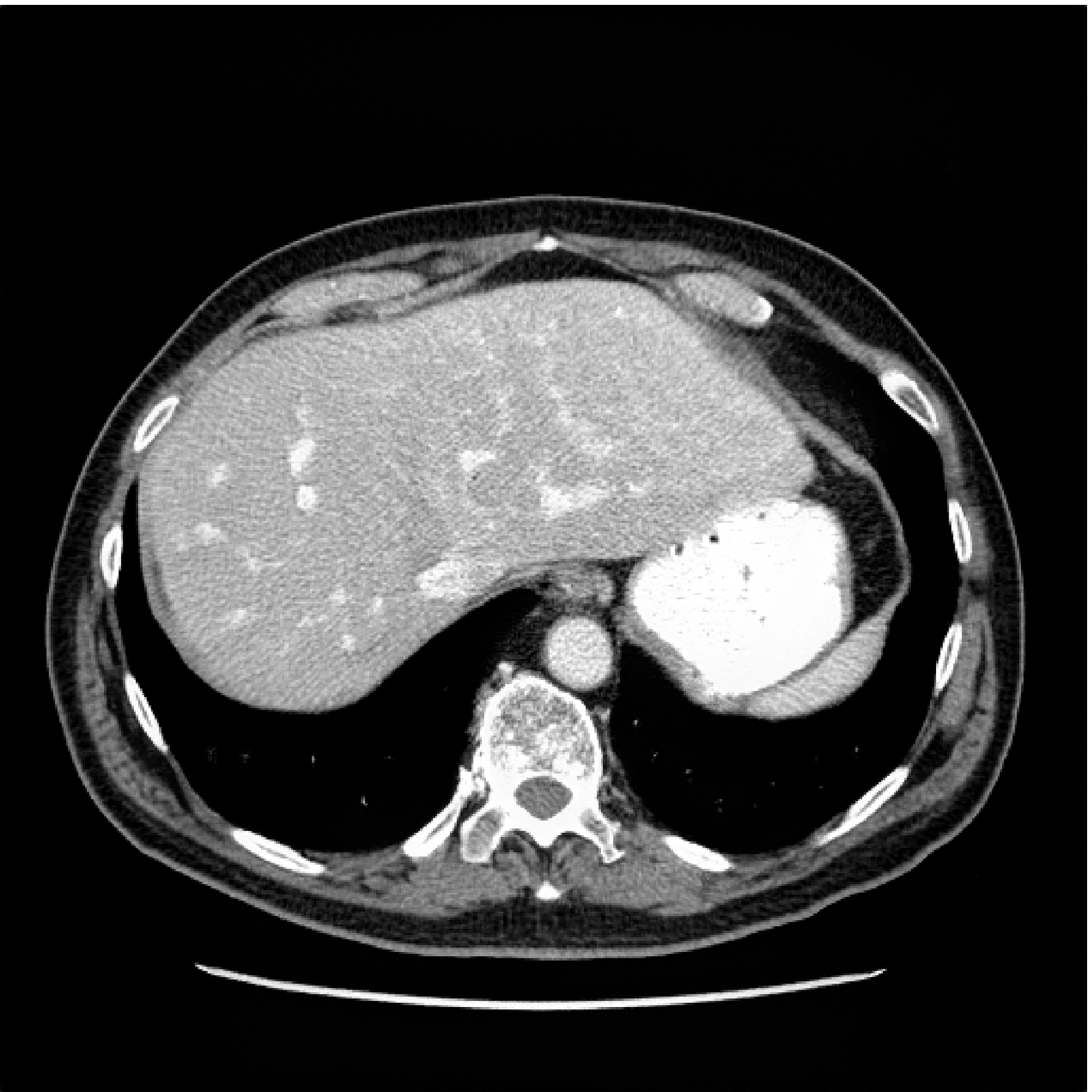}}
        \put(125,0){\includegraphics[width=0.27\textwidth]{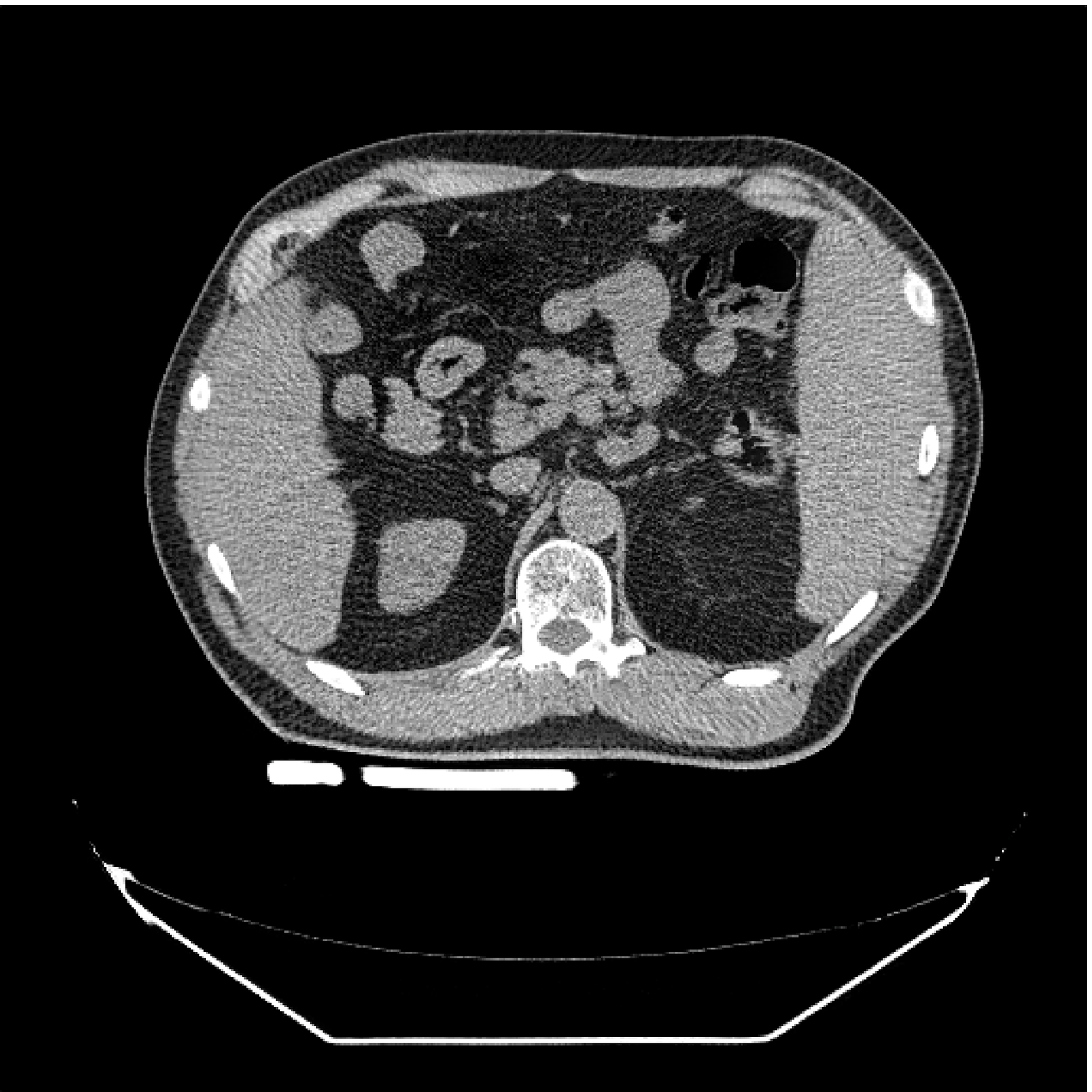}}
        \put(225,0){\includegraphics[width=0.27\textwidth]{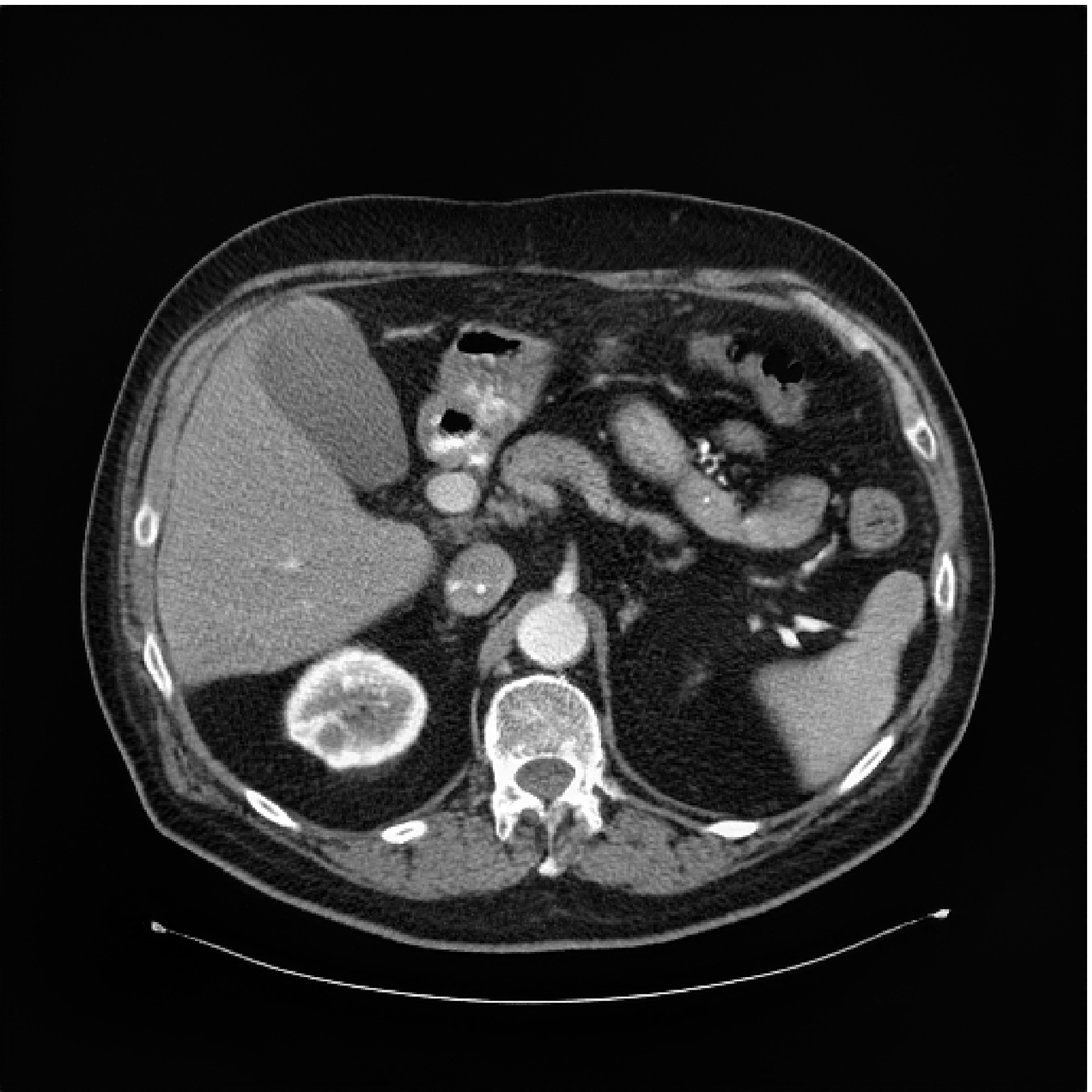}}
        \put(10,430){\rotatebox{90}{{\Large Real}}}
        \put(10,230){\rotatebox{90}{{\Large Baseline StyleGAN2}}}
        \put(10,7){\rotatebox{90}{{\Large Pretrained StyleGAN2-ADA}}}
    \end{picture}
    \caption{The first row contains images from our training dataset. 
    The second and third rows contain images generated by the baseline StyleGAN2 model.
    The fourth and fifth rows contain images generated by the pretrained StyleGAN2-ADA model with augmentations.
    All images were randomly selected.
    The images generated by the pretrained StyleGAN2-ADA model demonstrate reduced noise artifacts, enhanced detail, and superior anatomical accuracy}
    \label{fig:images}
\end{figure}

On our dataset, the average ($\pm$SD) FIDs, n=5, achieved were $10.70$ $(\pm 0.72)$, $7.62$ $(\pm 0.35)$, $7.51$ $(\pm 0.89)$, and $5.22$ $(\pm 0.17)$, for Experiments 1-4, respectively.
Both transfer learning and data augmentation were effective tools in mitigating overfitting on limited medical data.
Individually, they improved upon the baseline FID by about 30\% (95\% confidence).
Even greater improvements were achieved (50\% decrease) in the FID when transfer learning and augmentations were used in tandem (95\% confidence).
Data augmentation significantly decreased the generator's loss and stabilized training, as shown in Figure \ref{fig:loss} in the Appendix (95\% confidence).
Our results show that transfer learning does not need to be performed from a medical imaging dataset to be effective. When Experiment 1 was repeated with the additional 143,345 images, the average ($\pm$ SD) FID, n=5, attained was 8.45 $(\pm 0.20)$.
This demonstrates that transfer learning and data augmentation, both in conjunction and independently, outperformed a fifteenfold increase in the dataset size.

On the SLIVER07, ChestX-ray14, and ACDC datasets, we lowered the record FIDs from 29.06 to 10.78 (mean $11.99\pm 1.57$), 8.02 to 3.52 (mean $3.63\pm 0.07$), and 24.74 to 21.17 (mean $21.43\pm 0.32$), respectively.
For the Medical Segmentation Decathalon (brain tumors) data, we set a new record FID of 5.39 (mean $5.53\pm 0.01$).
These state-of-the-art results indicate that StyleGAN2 has stable performance and can generate quality medical images without a hyperparameter search.

Table \ref{tab:fid_fpr} shows the results of the multi-model visual Turing test.
This table provides empirical evidence that the FID is consistent with human perceptual judgement on medical images:
the lower the FID, the higher the average FPR (Pearson correlation of -0.91, 90\% confidence). This suggests that as the FID decreases, it becomes increasingly difficult for humans to distinguish between real and generated images.
In addition, the FPRs demonstrate that augmentations improved the perceptual quality of the generated images (90\% confidence).
When data augmentation was combined with transfer learning, the average participant was more likely to say a generated image was real than fake (55\% FPR).

\begin{table}
    \centering
    \caption{Average ($\pm$SD) results, n=5, of the multi-model visual Turing test. FIDs are associated with the model used to generate the images in the Turing tests.}
        \begin{tabular}{|l|c|c|c|}
            \multicolumn{4}{c}{\textbf{Multi-Model Visual Turing Test Results}} \\
            \hline
            \textbf{Experiment}
                &\textbf{FID}
                &\textbf{FPR}[\%]
                &\textbf{FNR}[\%]\\
            \hline
            1. Baseline 
                & 10.43
                & 29 ($\pm 27$)
                & 32 ($\pm 21$)\\
            \hline
            2. Pretrained
                & 7.78
                & 34 ($\pm 19$)
                & 32 ($\pm 18$)\\
            \hline
            3. Augmented
                & 7.15
                & 49 ($\pm 11$)
                & 34 ($\pm 18$)\\
            \hline
            4. Pretrained and Augmented
                &\textbf{5.06}
                &\textbf{55} ($\pm 9$)
                &\textbf{41} ($\pm 11$)\\
            \hline
        \end{tabular}

    \label{tab:fid_fpr}
\end{table}

\begin{figure}[h!]
    \centering
    \includegraphics[width=7cm]{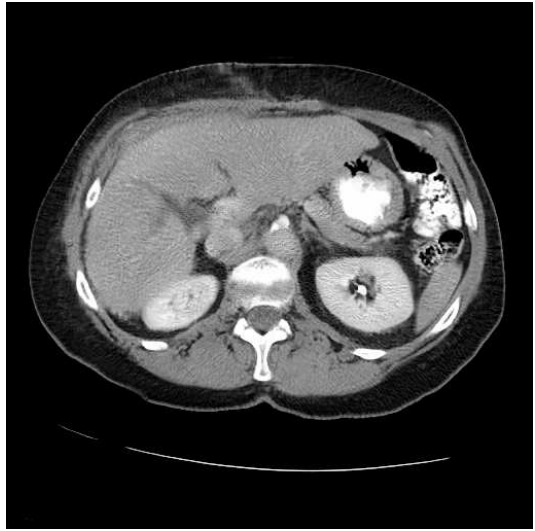}
    \caption{This image was generated by the baseline StyleGAN2 model (10.43 FID). It was chosen to demonstrate the noise artifacts contained in many of the images generated by the model.}
    \label{fig:noise}
\end{figure}

\begin{figure}[h!]
    \centering
    \begin{picture}(347,484)
        \put(0,360){\includegraphics[width=0.31\textwidth]{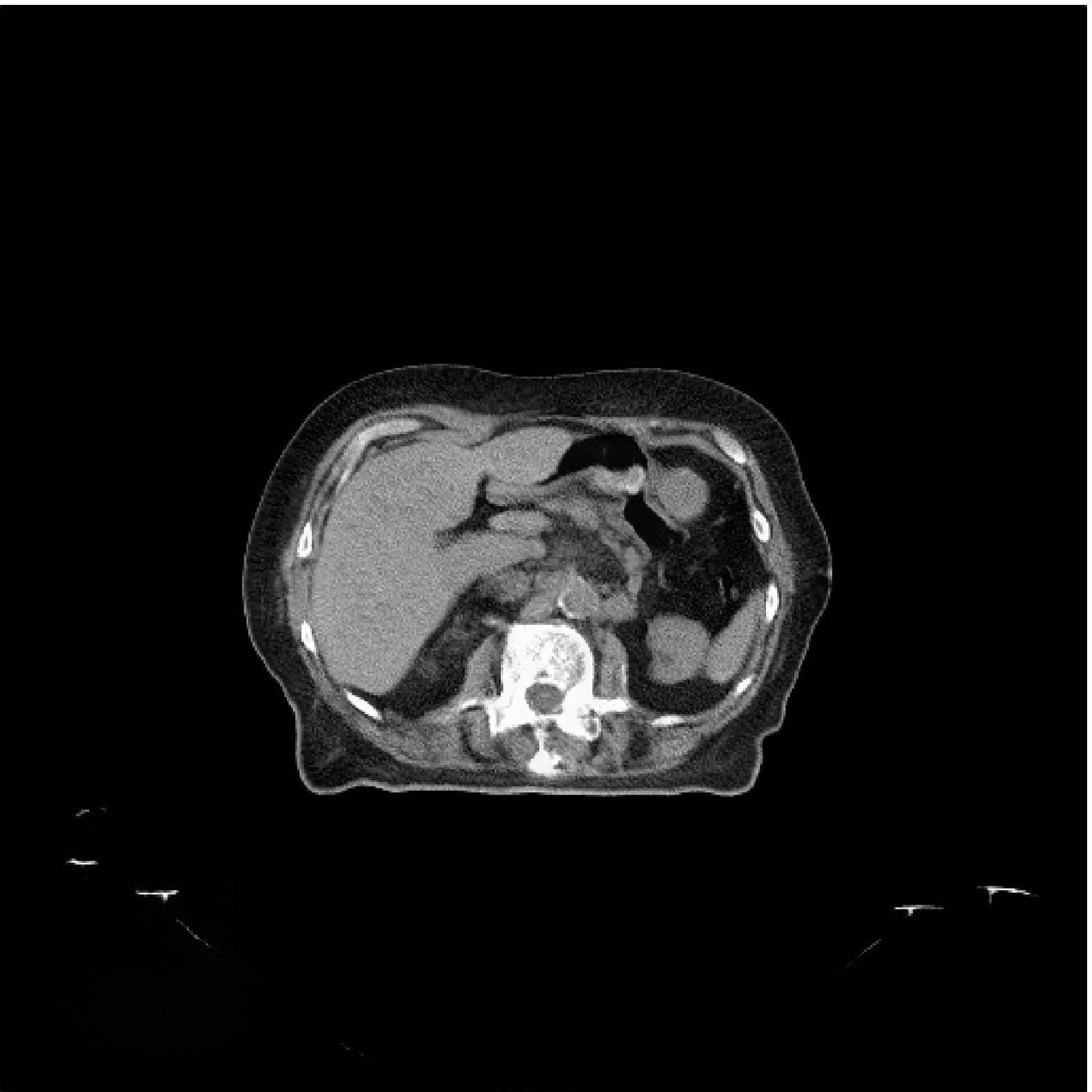}}
        \put(120,360){\includegraphics[width=0.31\textwidth]{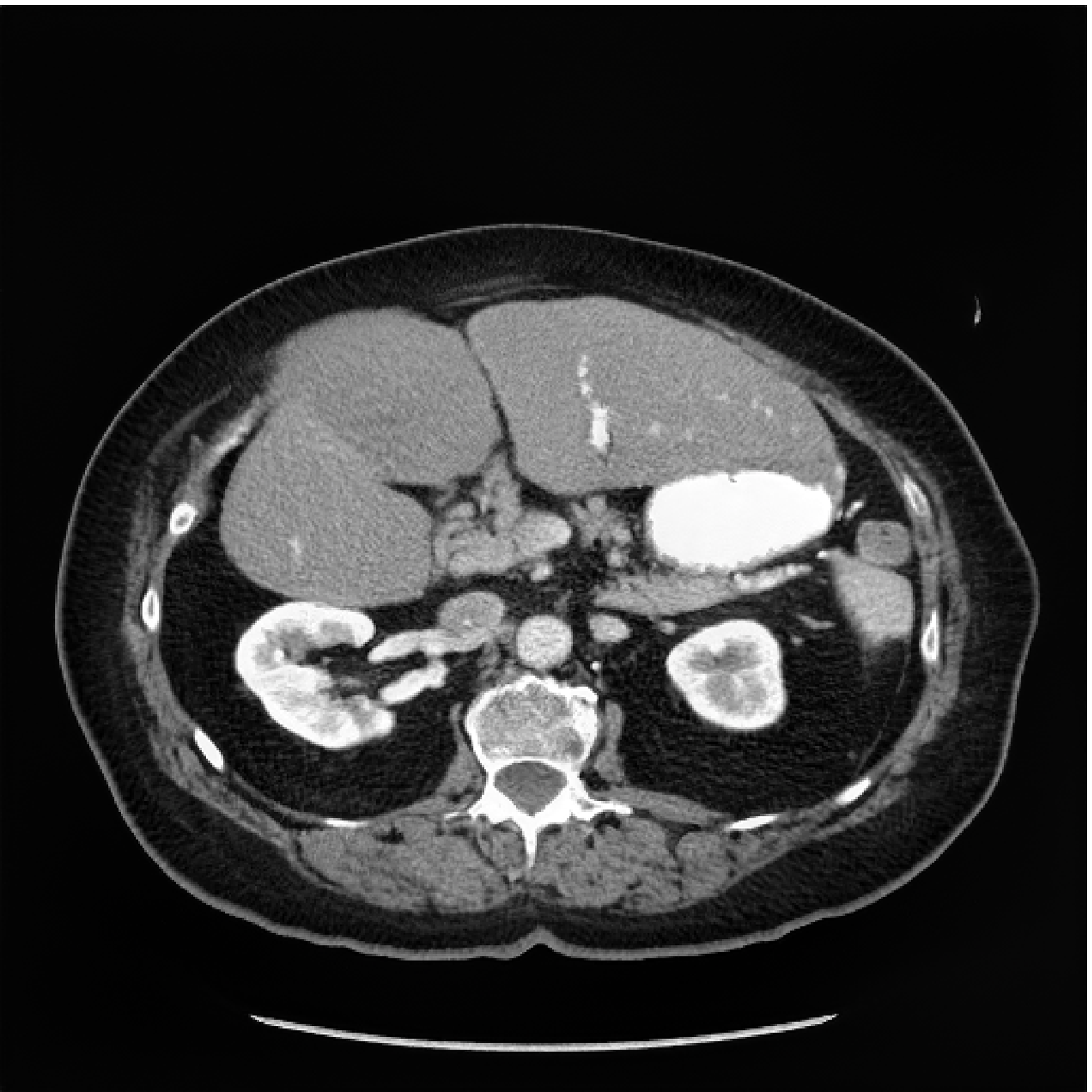}}
        \put(240,360){\includegraphics[width=0.31\textwidth]{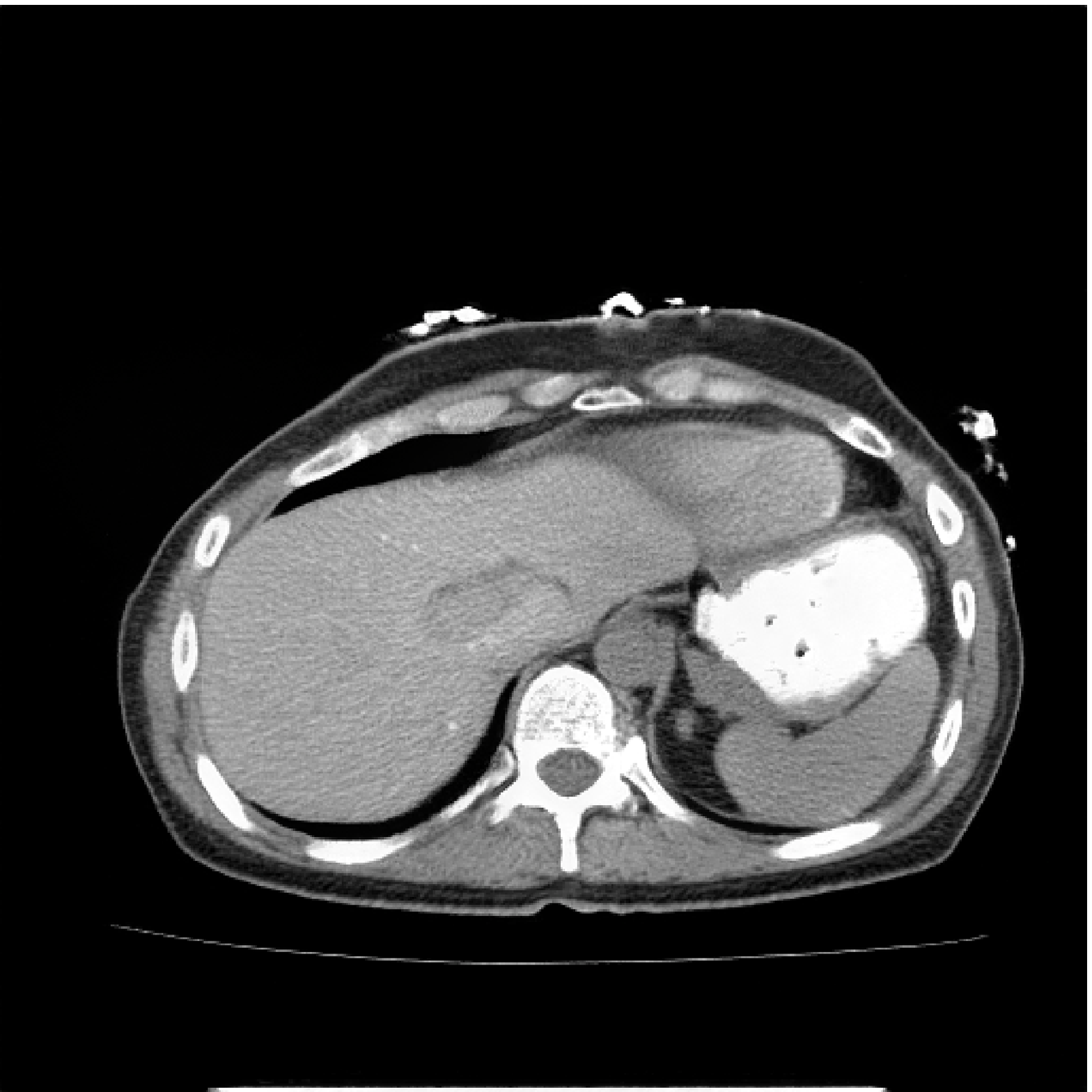}}
        \put(0,240){\includegraphics[width=0.31\textwidth]{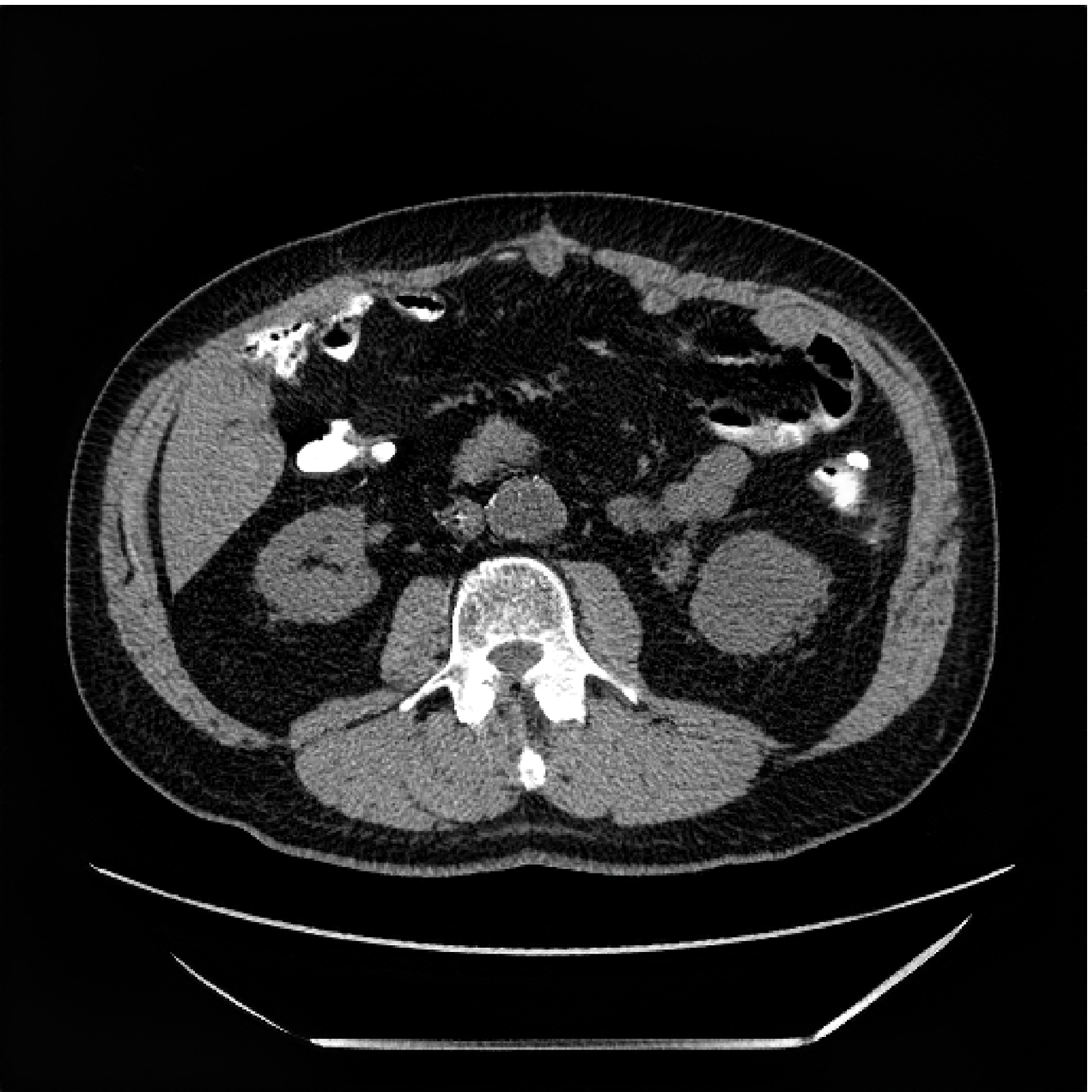}}
        \put(120,240){\includegraphics[width=0.31\textwidth]{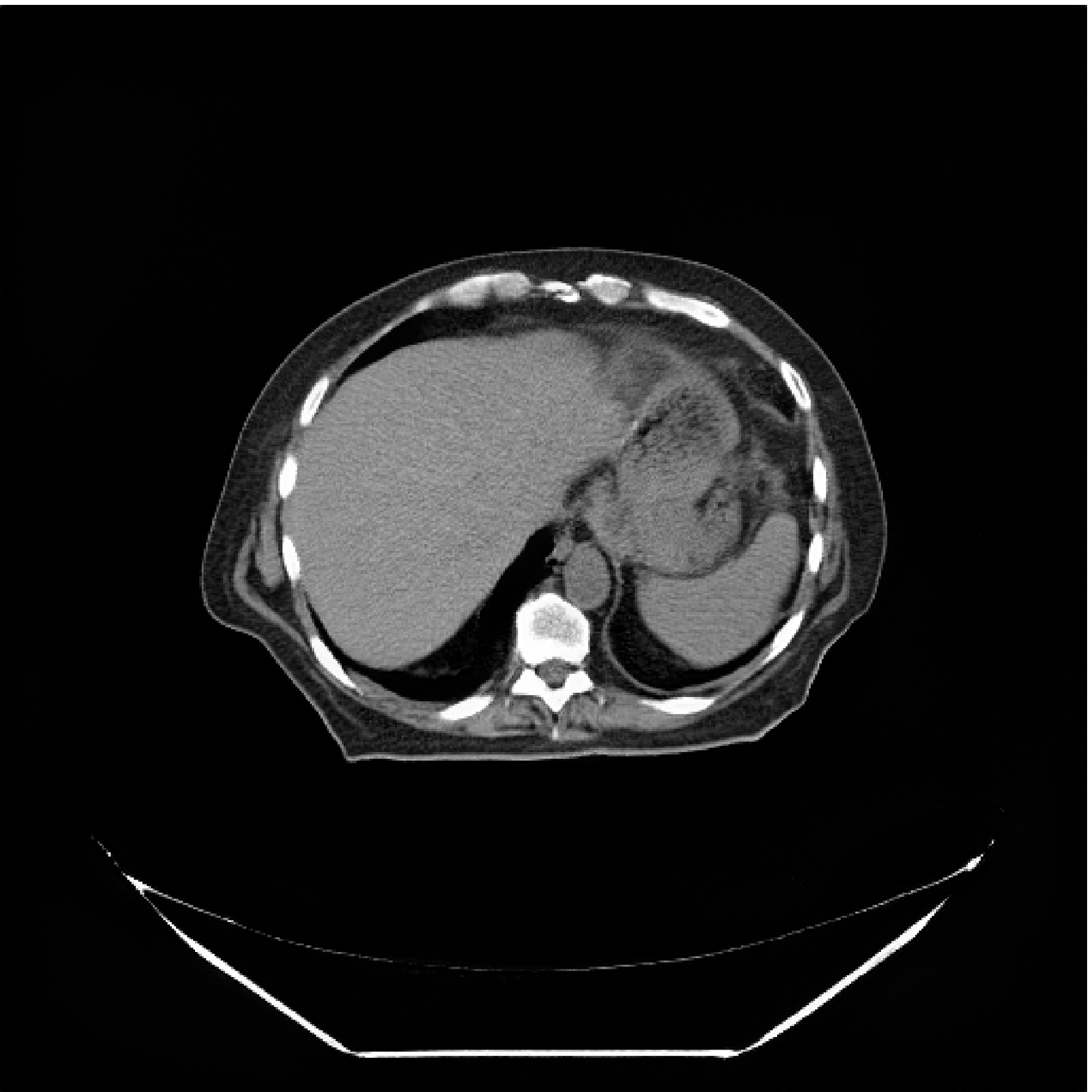}}
        \put(240,240){\includegraphics[width=0.31\textwidth]{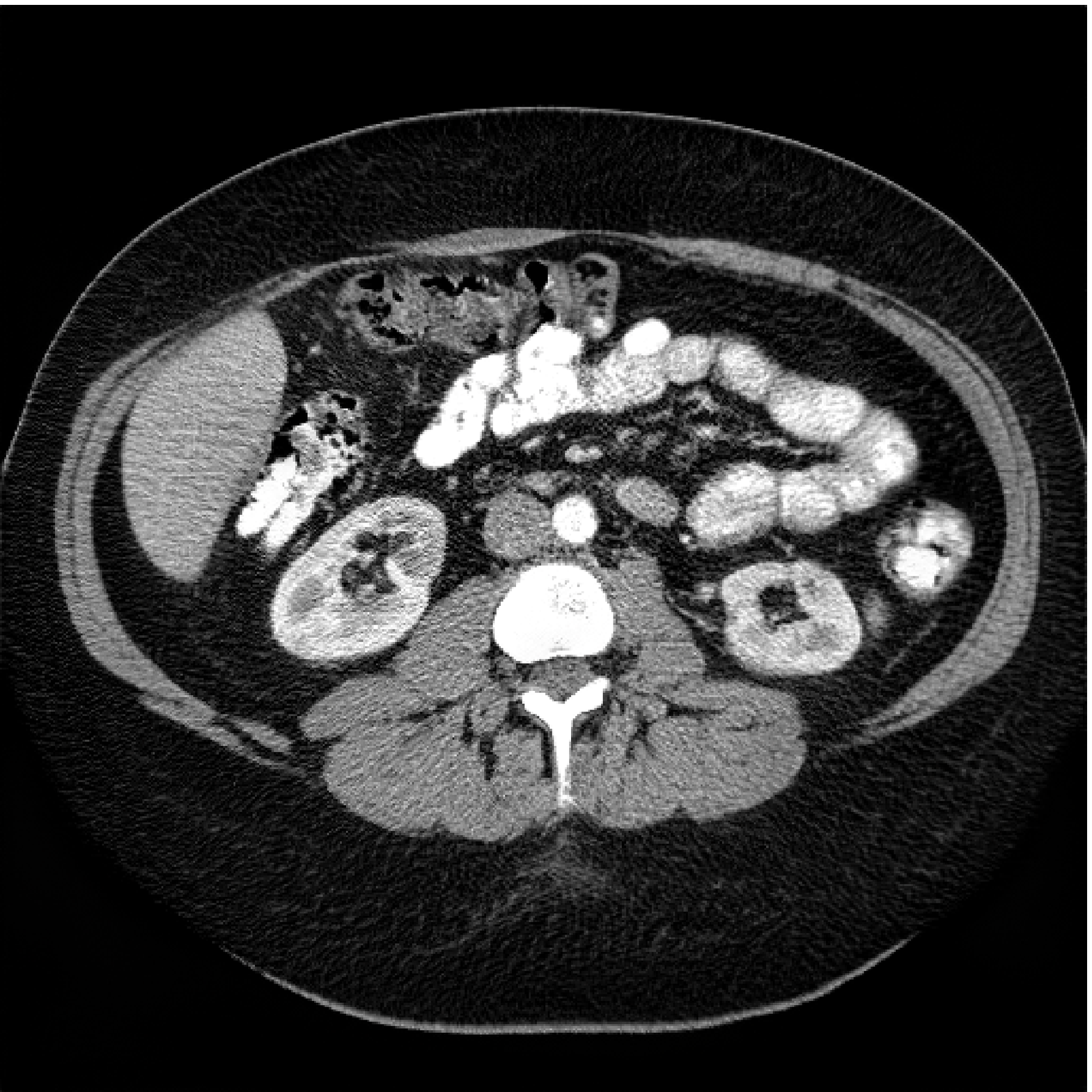}}
        
        \put(0,120){\includegraphics[width=0.31\textwidth]{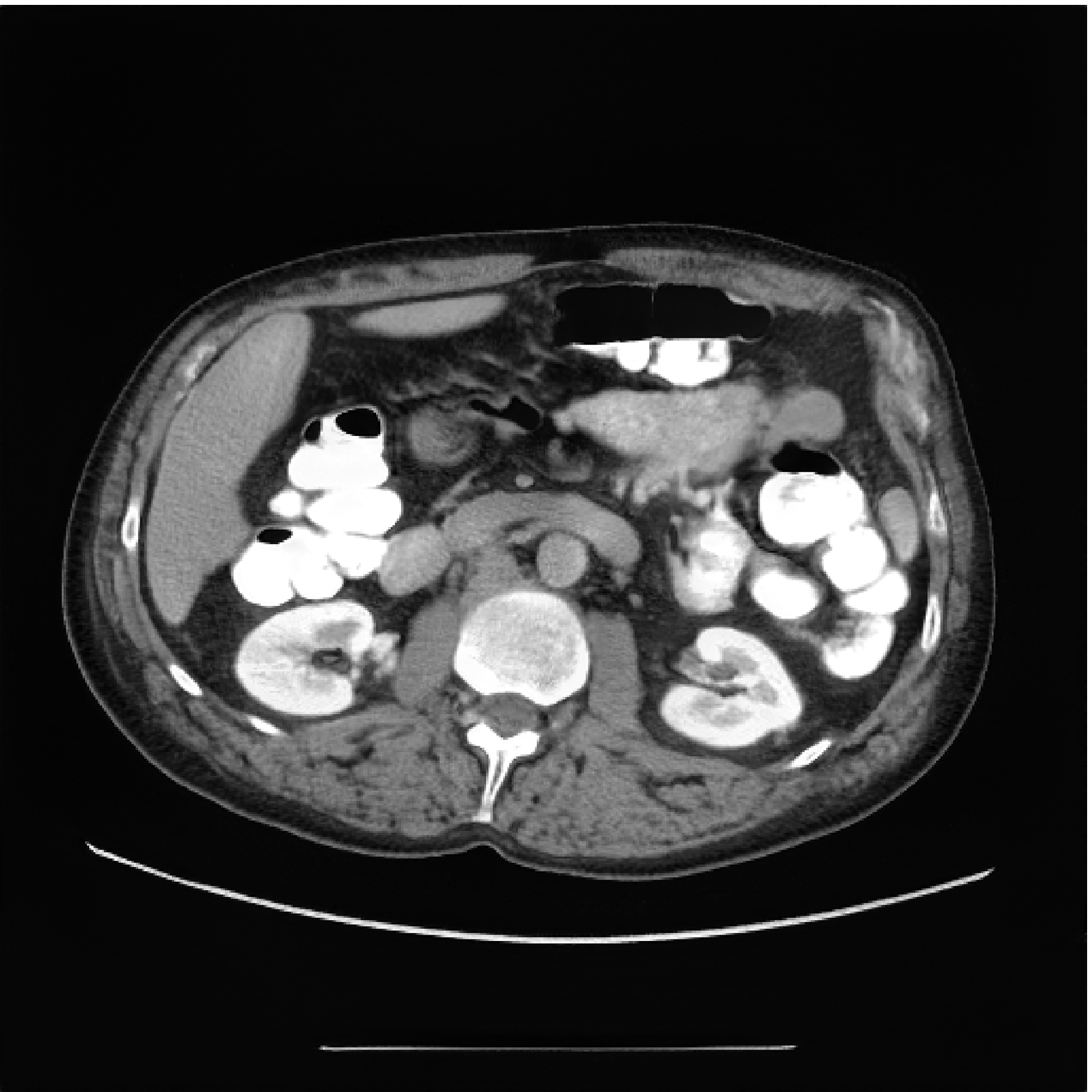}}
        \put(120,120){\includegraphics[width=0.31\textwidth]{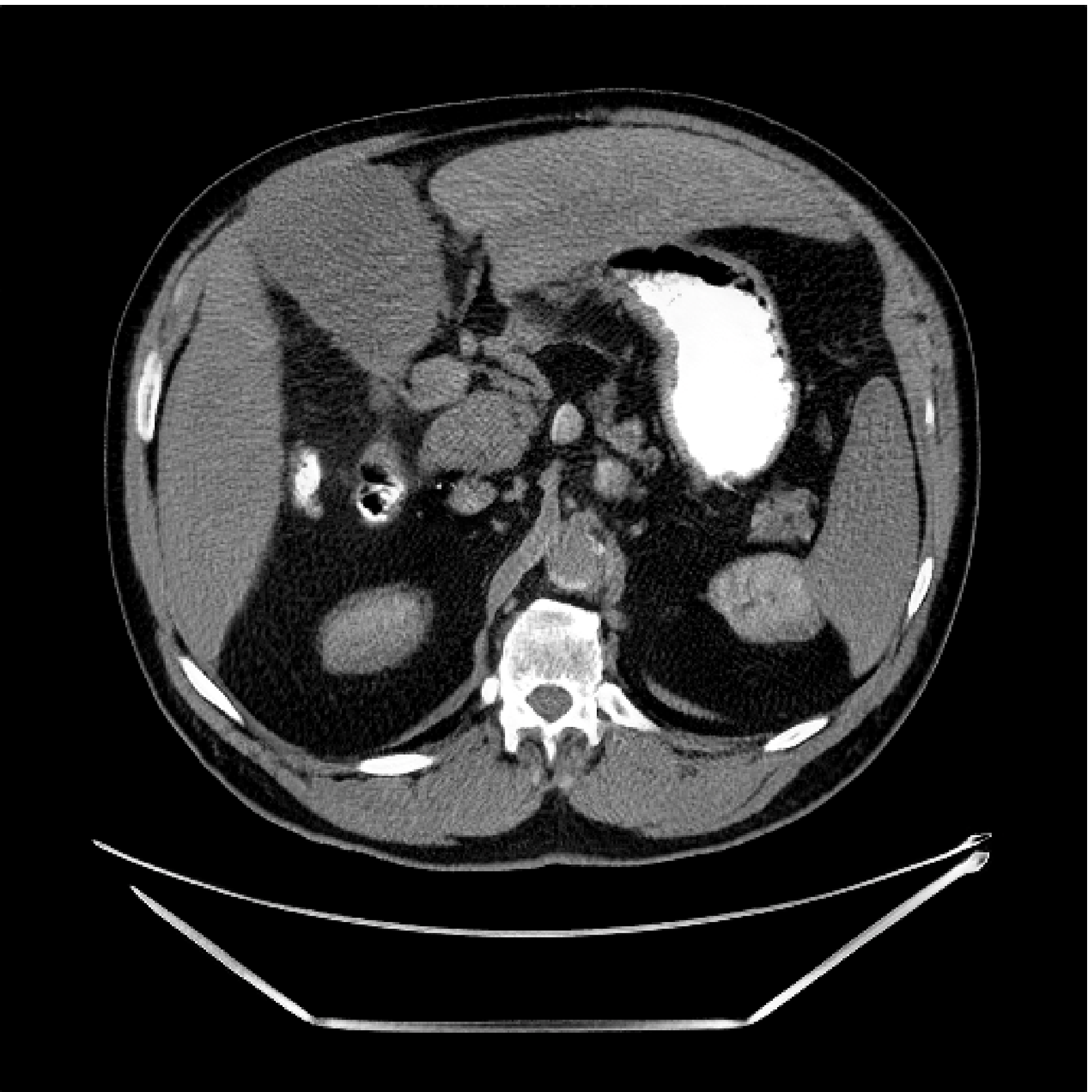}}
        \put(240,120){\includegraphics[width=0.31\textwidth]{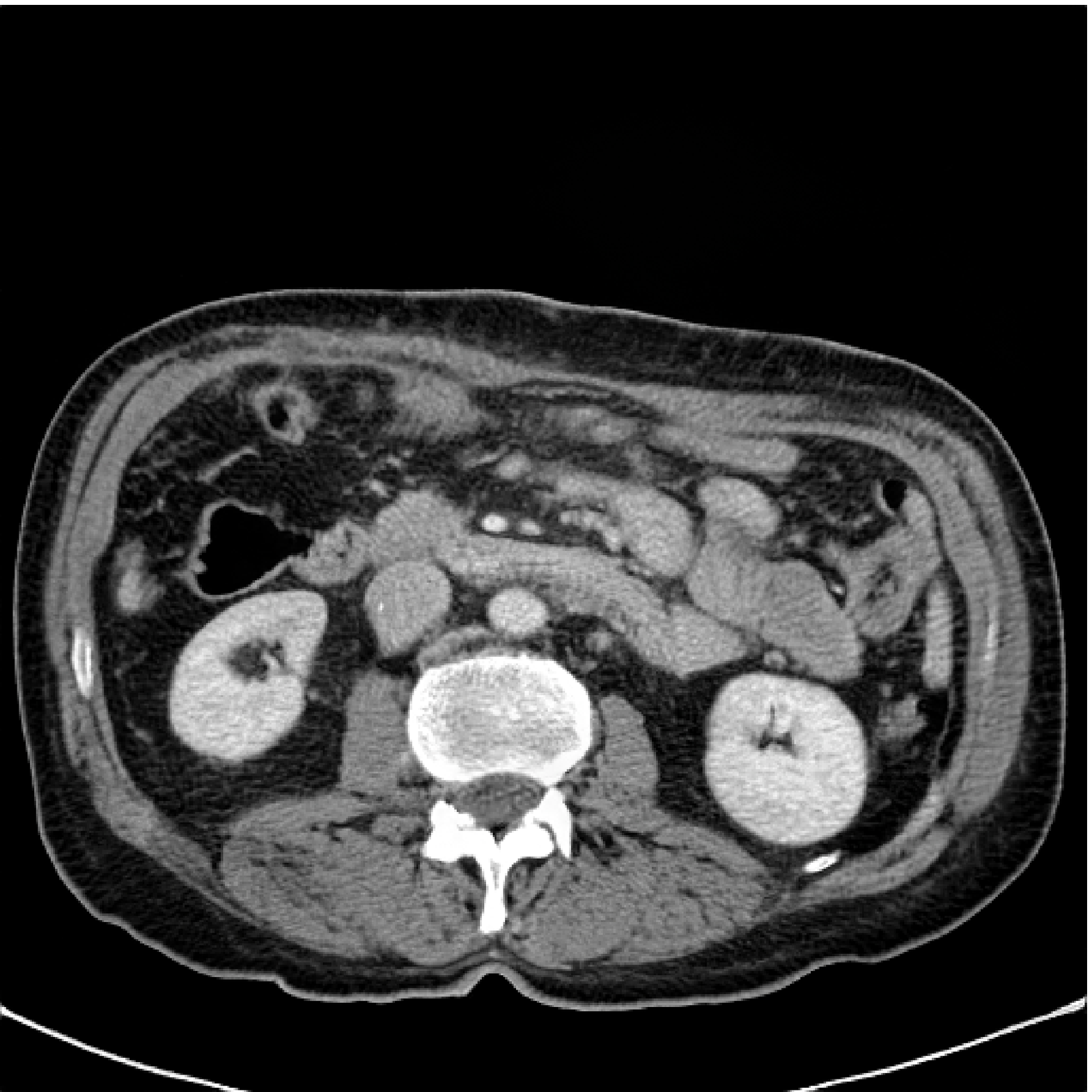}}
        
        \put(0,0){\includegraphics[width=0.31\textwidth]{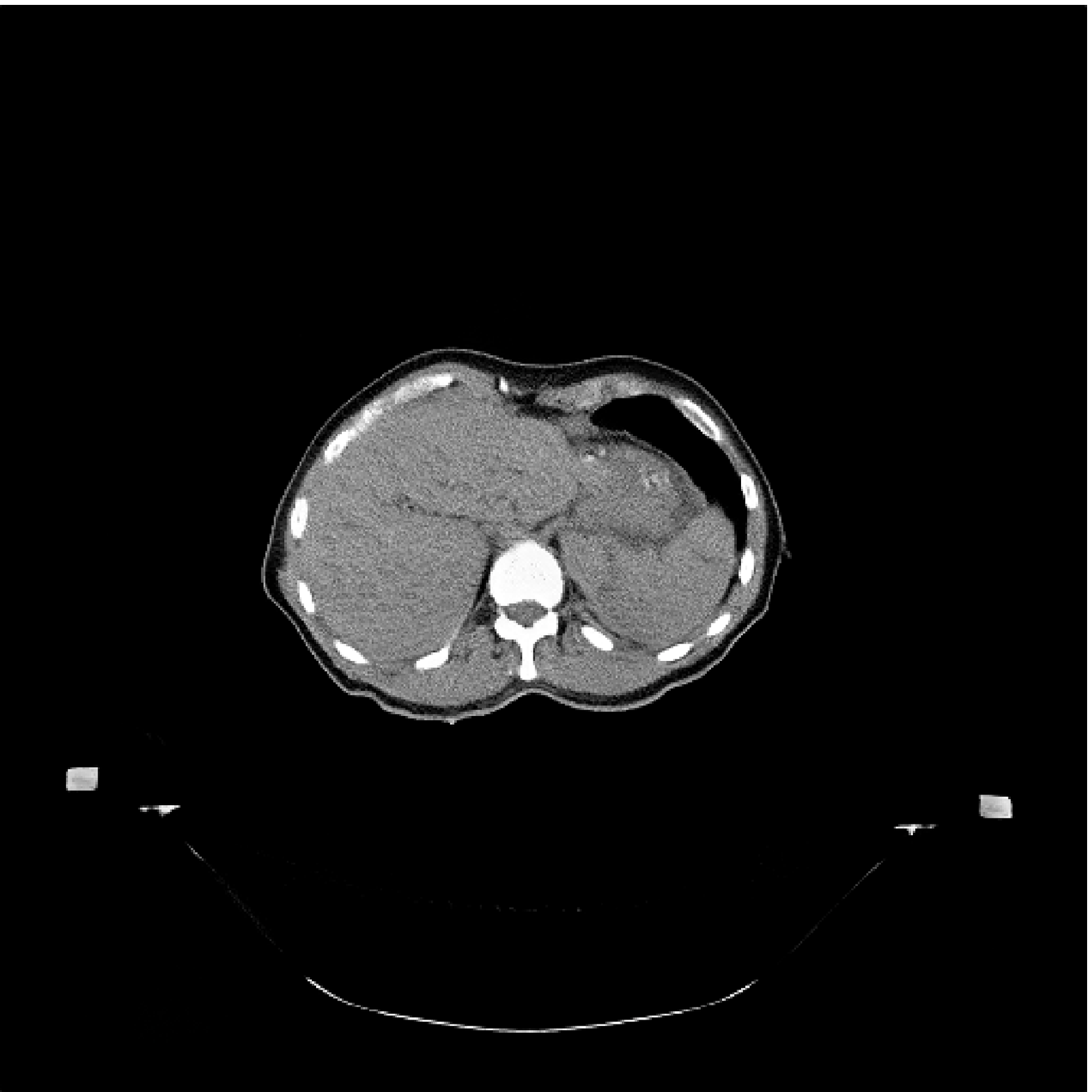}}
        \put(120,0){\includegraphics[width=0.31\textwidth]{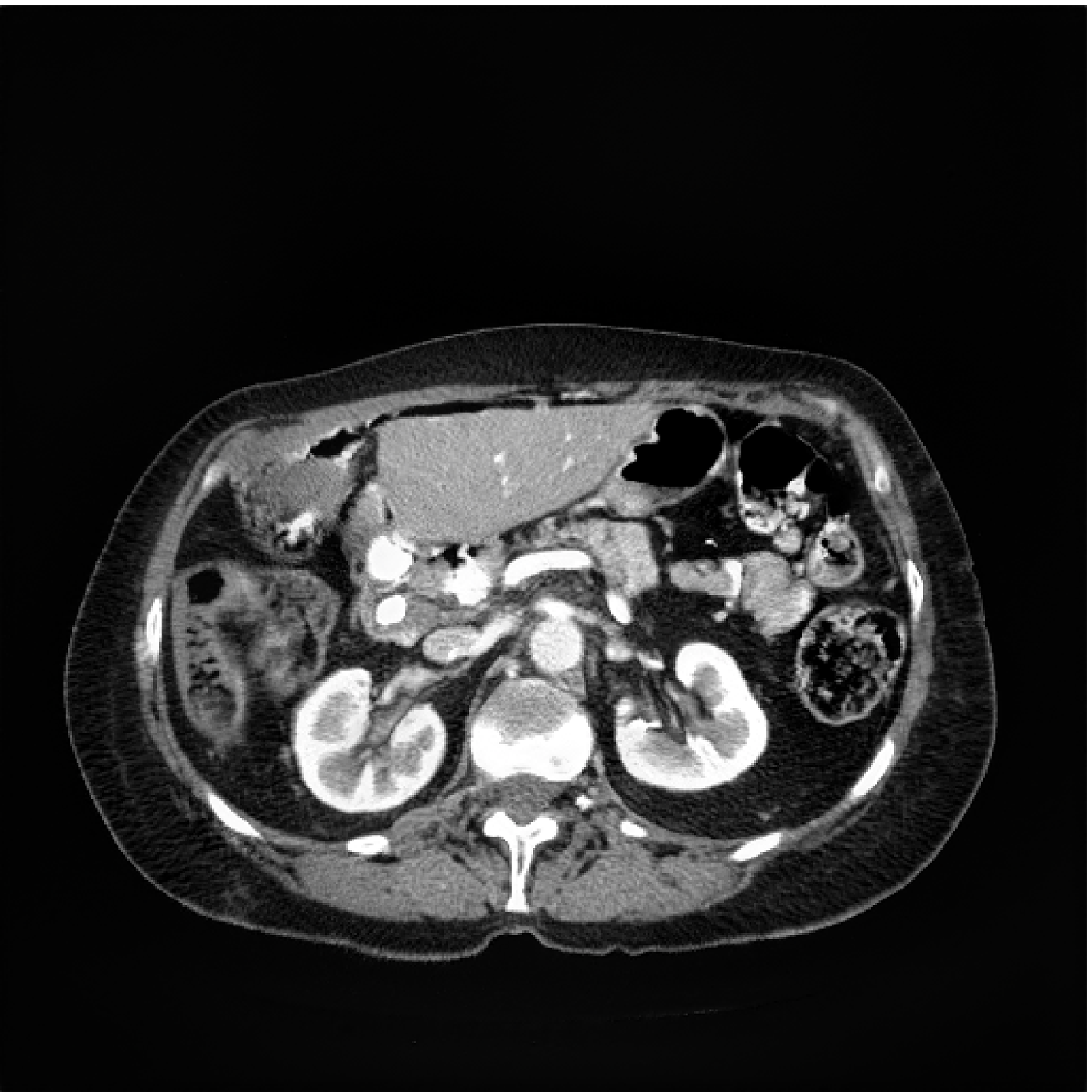}}
        \put(240,0){\includegraphics[width=0.3128\textwidth]{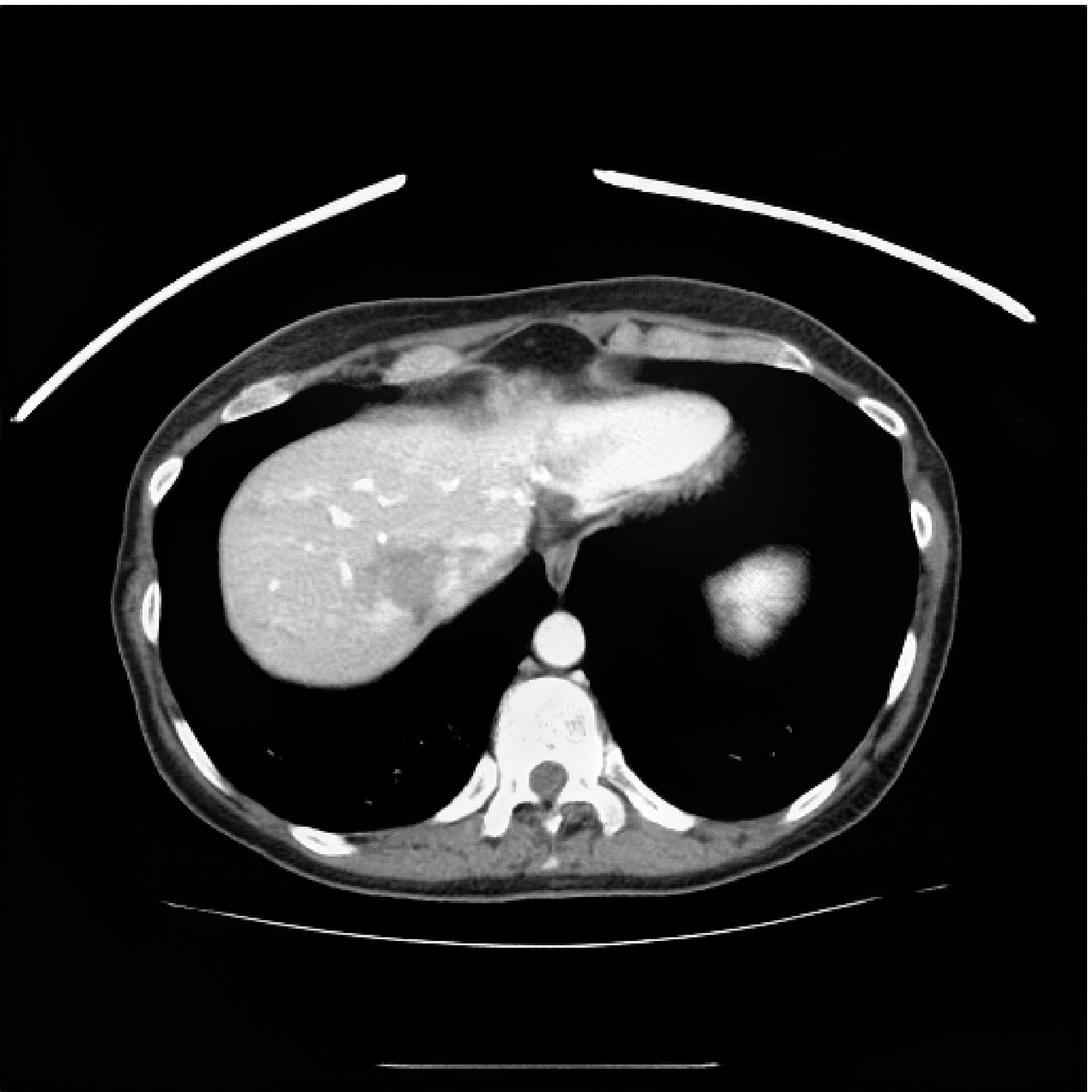}}
    \end{picture}
    \caption{Randomly selected images generated by the pretrained StyleGAN2-ADA model (5.06 FID).}
    \label{fig:generated}
\end{figure}

Figure \ref{fig:images} displays randomly selected real and generated images from the baseline StyleGAN2 (10.43 FID) and the pretrained StyleGAN2-ADA (5.06 FID) models.
Many of the images generated by the baseline StyleGAN2 model contain noise artifacts, especially in the liver.
Images generated by the pretrained StyleGAN2-ADA model show reduced noise, enhanced detail, and superior anatomical accuracy.
This perceptual improvement substantiates the claim that the FID is applicable to medical images.
The Appendix contains auxiliary pretrained StyleGAN2-ADA generated images (Figure \ref{fig:generated}) and a larger image demonstrating noise artifacts in the baseline StyleGAN2 model (Figure \ref{fig:noise}).

\begin{table}
    \centering
    \caption{Results of the visual Turing test given to clinicians.}
    \begin{tabular}{|c|c|c|c|c|c|c|c|c|}
        \multicolumn{6}{c}{\textbf{Clinician Visual Turing Test Results}} \\
        \hline
         \textbf{Clinician}
            & \textbf{Precision [\%]}
            & \textbf{Recall [\%]}
            & \textbf{Accuracy [\%]}
            & \textbf{FPR [\%]}
            & \textbf{FNR [\%]} \\
         \hline
        1 & 80 & 86 & 82 & 22 & 14\\
        \hline
        2 & 76 & 44 & 65 & 14 & 56 \\
        \hline
        3 & 56 & 80 & 58 & 64 & 20 \\
        \hline
        4 & 79 & 62 & 73 & 16 & 38 \\
        \hline
        5 & 58 & 98 & 64 & 70 & 2 \\
        \hline
        6 & 54 & 88 & 56 & 76 & 12 \\
        \hline
        7 & 59 & 48 & 57 & 34 & 52 \\
        \hline
        \textbf{Average} ($\pm$SD)
            & 66 ($\pm 12$)
            & 72 ($\pm 21$)
            & 65 ($\pm 10$)
            & 42 ($\pm 27$)
            & 28 ($\pm 21$)\\
        \hline
    \end{tabular}
    \label{tab:vtt}
\end{table}

The results of the Turing test given to clinicians, shown in Table \ref{tab:vtt}, further confirm the high-quality nature of the generated images.
Overall, the clinicians classified generated images as real 42\% of the time, approaching the equivalent of random guessing.
Those that had low FPRs typically had higher FNRs and vice versa (Pearson correlation of -0.71, 90\% confidence), indicating a tendency of clinicians to favor either ``real'' or ``fake'' when they were unsure.
This tendency was likely a factor in the high interobserver variability among the FPRs.
Another likely factor was the experience of the clinicians.
For the Likert scale, we found that real images achieved an average score of 3.99 ($\pm$ 1.00) and generated images a score of 3.23 ($\pm$ 1.21).
The overlapping 95\% confidence intervals further demonstrate both the challenging nature of the task and the high-quality nature of the generated images.

\section{Conclusion}

We applied StyleGAN2 to multiple high-resolution medical image datasets.
Combined with transfer learning and data augmentation, the architecture achieved state-of-the-art results consistently, without any hyperparameter searches or retraining.
The generated images were of sufficient quality that an expert's ability to tell whether or not an image was generated approached random guessing.
Additionally, we found that the ``realness'' score, based on a 5-point Likert scale, differed between the generated and real images by less than the standard deviation between clinicians.
Across a variety of medical imaging modalities, we were able to set new record FID scores on four publicly-available datasets.

Furthermore, our research provided empirical evidence that the FID is consistent with human perceptual judgement on medical images.
A multi-model visual Turing test revealed that as the FID improved, the participants perceived artificially generated images as real more frequently.
Qualitatively, we saw an appreciable improvement in the fidelity of the generated images as the FID improved from 10.43 to 5.06.
From these results, we concluded that the FID is indeed an appropriate metric for medical images.

\subsubsection{Acknowledgements}

This work was supported by the Tumor Measurement Initiative through the MD Anderson Strategic Initiative Development Program (STRIDE). 
We thank the NIH Clinical Center for the ChestX-ray14 dataset.

\clearpage

\section*{Appendix}

\begin{figure}[h!]
    \centering
    \includegraphics[width=7cm]{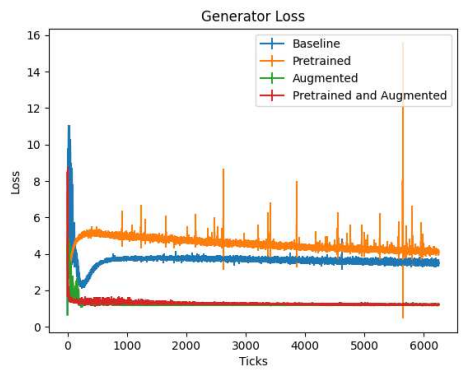}
    \caption{The average generator loss (with standard deviation bars) across training.
    We see that augmentation not only significantly decreases the loss, but also leads to more stable convergence.}
    \label{fig:loss}
\end{figure}

\clearpage

%
%
%
\bibliographystyle{splncs04}
\bibliography{ref}
\end{document}